\documentclass[%
 aip,
 amsmath,amssymb,
preprint,%
]{revtex4-2}

\usepackage{graphicx}
\usepackage{subfigure}
\usepackage{dcolumn}
\usepackage{bm}
\usepackage{soul}
\usepackage[dvipsnames]{xcolor}

\usepackage[utf8]{inputenc}
\usepackage[T1]{fontenc}
\usepackage{mathptmx}
\usepackage{etoolbox}

\usepackage{xcolor}
\renewcommand*{\thefootnote}{\fnsymbol{footnote}}

\makeatletter
\def\@email#1#2{%
 \endgroup
 \patchcmd{\titleblock@produce}
  {\frontmatter@RRAPformat}
  {\frontmatter@RRAPformat{\produce@RRAP{*#1\href{mailto:#2}{#2}}}\frontmatter@RRAPformat}
  {}{}
}%

\makeatother
\begin{document}

\preprint{AIP/123-QED}

\title[Shock train - shear layer interaction]{Interaction of Shock Train with Cavity Shear Layer in a Scramjet Isolator}
\author{Vignesh Ram Petha Sethuraman}
\affiliation{ 
Department of Aerospace System Engineering, Sejong University, Seoul, South Korea - 05006.
}
\author{Yosheph Yang}
\affiliation{%
Department of Aerospace Engineering, Sejong University, Seoul, South Korea - 05006.
}
\author{Jae Gang Kim*}%
 \email{jaegkim@sejong.ac.kr}
\affiliation{ 
Department of Aerospace System Engineering, Sejong University, Seoul, South Korea - 05006.
}%

\date{\today}

\begin{abstract}
The interaction between the self-excited shock train flow and the cavity shear layer in a scramjet isolator is investigated numerically using detached-eddy simulations (DES). The effect of changing the position of the shock train by controlling the back pressure ratio and the effect of changing the cavity front wall angle are analyzed using unsteady statistics and modal analysis. The propagation mechanism of the pressure disturbance was investigated by spatiotemporal cross-correlation coefficient analysis. In the present numerical study, a constant isolator section with a cavity front wall ($\theta$ = 90$^o$ and 60$^o$) was considered, followed by a diffuser section simulated at Mach number 2.2 with three different back pressure ratios (p$_b$/p$_{\infty}$ = 0.7, 5.0, and 6.0). The change in back pressure provides three different conditions (i.e., no shock train, shock train ends before the leading edge of the cavity, and shock train present above the cavity). To understand the unsteady dynamics of the interaction of the shear layer with the shock train, the spatiotemporal trajectory of the wall pressure and the centerline pressure distribution, the spatiotemporal cross-correlation coefficient, and the modal analysis by dynamic mode decomposition are obtained. The results show that the low-frequency shock train oscillation dominates the self-sustained cavity oscillation. The spatiotemporal cross-correlation between the wall surface and the center of the cavity bottom wall indicates the propagation of local disturbances originating from the separated boundary layer caused by the shock and the recirculation zone in the corners of the cavity. Dynamic mode decomposition analysis shows the shear layer at the leading edge of the cavity and the downstream propagation of large eddies from the cavity. It also shows the pairing of coherent structures between the shock train and the recirculation zone of the cavity.
\end{abstract}

\maketitle

\section{\label{sec:label1}INTRODUCTION:\protect}
 
  The presence of shock waves in high-speed internal flow devices such as the supersonic combustion ramjet (scramjet) is common and unavoidable. The scramjet engine consists of four main parts, namely an inlet, an isolator, the combustion chamber, and a nozzle. The compression process occurs through various types of shock waves such as oblique, incident, and reflected shock structures that occur at the inlet. When a scramjet is operated under supercritical flow conditions, the normal shock usually occurs in the isolator, which is also referred to as terminal shock. Interacting with the incoming boundary layer, the terminal shock undergoes three shock patterns (i.e., a curved shock wave, a lambda shock wave, and a shock train). In the case of the curved shock wave and the lambda shock wave, the interaction between the normal shock and the boundary layer is quite weak. When the interaction becomes too strong, a single normal shock bifurcates into several successive normal shock structures known as a "shock train". If the isolator section is long enough, the static pressure continues to increase until the frictional force dominates. This region is called the "mixing region." The formation of the shock train in a constant area channel and its properties are described in detail by Matsuo et al. \cite{matsuo_1999a}. 

  Unlike normal shocks, the length of the shock train is usually on the order of several duct heights. For this reason, the isolator section is designed to accommodate the entire shock train system. The isolator provides sufficient static pressure in the combustion system to allow an efficient combustion process through the shock train system. At the same time, this combustion process not only causes the downstream static pressure to increase but also causes the entire shock train system to move upstream toward the intake area, resulting in engine unstart condition \cite{Sullins_1993, Heiser_1994,wagner_2009,chang20171,devaraj_2020} or inlet buzz condition \cite{im_2018, rajasekar_2020, jintu_2021, zhang_2022}. The steady-state flow characteristics of the shock train depend mainly on the pressure ratio (p$_2$/p$_1$), inlet freestream Mach number (M$_{\infty}$), the aspect ratio of the duct (AR = L/H), and the blockage ratio ($\beta$) \cite{carroll_1990, om_1985b, tamaki_1970,tamaki_1971}. These parameters determine the position of the shock train, its length, the number of successive shocks, and its structures. In recent decades, the steady-state properties of the shock train, such as its position, length, and static pressure rise, have been estimated using various one-dimensional models \cite{ikui_1974a,ikui_1974b,matsuo_1999b} and empirical relations \cite{waltrup_1973,weiss_2010, geerts_2016, ram_2020}. However, shock train flows exhibit unsteadiness due to the separated boundary layer and downstream pressure perturbations \cite{ikui_1974b}. Such oscillatory flows can be classified as high-amplitude shock wave oscillations. In contrast, the low-amplitude shock wave oscillations are related to the induced shock wave over a compression ramp or the oblique incident shock wave \cite{delery_2009}. 

  Studies related to shock train unsteadiness have recently attracted increasing attention from researchers. In the case of a uniform, unperturbed inlet flow, the shock train oscillates in phase with the frequency of the downstream pressure perturbation and behaves like a second-order non-homogeneous differential equation \cite{ram_2021}. Varying the incoming flow properties \cite{nanli_2017, nanli_2018, shi_2019a, shi_2019b, hou_2020a, wang_2020, lihao_2022, gang_2022} and the downstream back pressure \cite{xiong_2018a, saravanan_2020, wang_2021, wangyi_2021, wang_2022} with background waves results in different types of shock train oscillations that have a greater impact on the performance of the scramjet. The low-frequency oscillations of shock train initiate asymmetric flowfield and lead to an intrinsic istability\cite{chen_2019}. Various active and passive control methods are used to control such low-frequency high-amplitude oscillations of shock train\cite{su_2018, Meng_2020, nanli_2022}. One of the active control methods, such as partial removal boundary layer flow through the suction slot, not only helps to alleviate the shock wave oscillations but also reduces the length of the shock train \cite{weiss_2012, ram_2021b,ram_2022}. Moreover, the shock train oscillates at the same frequency but out of phase with the flame oscillation \cite{micka_2009}. A major challenge faced by a dual-mode scramjet engine is the stabilization of combustion (i.e., auto-ignition and fuel mixing), which can be achieved by using a cavity downstream of the fuel injection. Although the cavity flame holder has the advantage of reducing the total pressure loss, aerodynamic heating \cite{fotia_2013, gruber_2001, gruber_2015, micka_2009} and sudden transition between ram to scram mode\cite{ruixu_2022}. The mechanism of combustion stabilization and mode transistion phenomenon in high-speed flows is not fully understood. Before combustion stabilization can be studied, the mechanism of interaction between the pre-combustion shock train and the cavity shear layer must be understood. Both the self-sustaining oscillation of the shock train and the oscillation of the cavity flow are completely different mechanisms. Self-sustained oscillations of the shock train are triggered by a small perturbation of the downstream flow or the local boundary layer flow. On the other hand, the interaction between the shear layer and acoustic disturbances from the recirculation zone leads to self-sustained oscillatory flow in the cavity \cite{sarohia_1977, tam_1978, heller_1996, woo_2008}. Extensive experimental and numerical studies on cavity flow oscillation and its control techniques can be found in the open literature \cite{youngkilee_2008, turpin_2020, aravind_2019}. Most of these studies assume that the flow is free of shock interactions. However, in scramjet flow, the cavity is subjected to various types of shock interactions, such as incident shocks, and bow shocks that appeared due to fuel injection \cite{javier_2018}, resulting in a more complex oscillation mechanism. Cheng et al \cite{chengkuo_1998} conducted experimental studies on the interaction of the induced shock with the shear layer of the cavity and found that the incident shock alters the oscillatory behavior by reducing the local momentum thickness and enhancing the acoustic feedback mechanism. Recent numerical studies on the interaction of wedge-induced oblique shocks with the shear layer of the cavity show an increase in static wall pressure of about 25\% compared to the case without interaction \cite{karthick_2021}. Studies on the interaction between normal and oblique shock waves and the shear layer in cavities are limited and mostly deal with the interaction of low-amplitude shocks with the cavity shear layer. As mentioned earlier, shock train flows are high-amplitude flows that can oscillate on the order of the duct height. Therefore, the cavity flows experience a repeated supersonic-subsonic flow condition with shock train oscillation. Therefore, in the present work, the interaction between the shock train and a cavity shear layer in an isolator is studied with the following objectives:
  
  \begin{enumerate}
    \item To numerically simulate the shock train flows in the proximity of wall mounted cavity in an isolator.
    \item To compare the transient characteristics of shock train–shear layer interaction with different back pressure ratios and cavity front wall angle.
    \item To investigate the power spectrum along the flow field and identification of common features between the different cases.
    \item To analyze the mechanism of local disturbance propagation using the two-point cross-correlation method.  
    \item To identify the dominant spatiotemporal modes using the dynamic mode decomposition (DMD) method.
  \end{enumerate}

  The outcome of the present work can help to understand the effects of the interaction of the shock train with the cavity shear flow and its behavior. The paper is organized as follows. In Sec.\ref{sec:2}, the problem is briefly introduced with the flow field theory ( Sec.\ref{sec:2a}), geometrical details (Sec.\ref{sec:2b}), and flow configuration (Sec.\ref{sec:2c}). The numerical procedure, grid independence study, and reliability of the unsteady simulation are described in Sec.\ref{sec:3}. The steady and unsteady statistics, the spectral analysis of the shock train shear layer interactions, the mechanism of the disturbance propagation, and the modal analysis are described in Sec.\ref{sec:4}. In the last section (Sec.\ref{sec:5}), the major outcomes of the present study are narrated.
  
\section{PROBLEM STATEMENT}
\label{sec:2}
  To compare the dynamics of the shock train and shear layer interaction, a problem statement has been constructed using the numerical study of the scramjet isolator along with a cavity (see Fig.1). A terminal shock in the scramjet isolator interacts with the boundary layer and forms a shock train (at M$_{\infty}$ $>$ 1.5) \cite{matsuo_1999a}. The increase in back pressure leads to an upstream movement of the shock train. The current study focuses on the shock train present upstream of the cavity (similar to the ramjet mode) and the shock train present above the cavity location (similar to the scramjet mode). Two different variations of the angle of the cavity front wall are used ($\theta$ = 90$^o$ and 60$^o$), as shown in Fig.2. At p$_b$/p$_{\infty}$ = 0.7, the cavity experiences complete supersonic flow with no shock train in the isolator section. At p$_b$/p$_{\infty}$ = 5.0 and 6.0, the shock train is located near and upstream of the cavity, respectively. The present cases were simulated as cold flow without combustion or fuel injection. A brief description of the interaction between the shock train and the cavity shear layer is discussed in Sec.\ref{sec:2a}.
  
  \begin{figure}
  \includegraphics[width=0.7\linewidth]{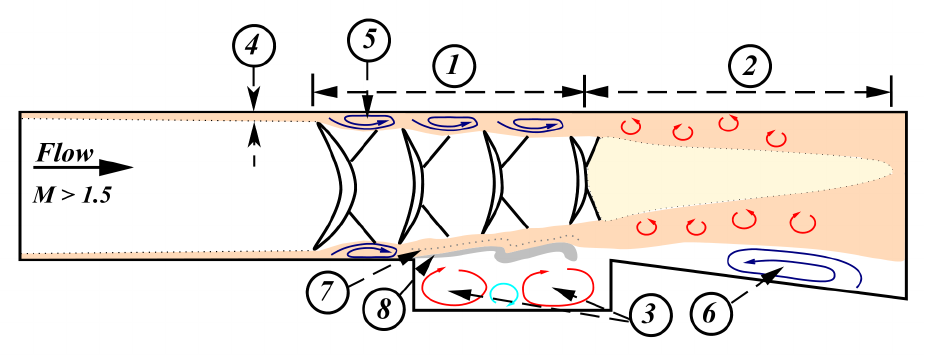}
  \caption{\label{fig:shock_shear_interaction} A simple schematic of shock train - cavity shear layer interaction. (1) Shock train region (2) mixing region (3) cavity recirculation region (4) incoming flow boundary layer thickness (5) shock train leading shock separation region (6) diffuser section flow separation region (7) shear layer due to separated boundary layer of shock train region (8) cavity shear layer}
  \end{figure}
\subsection{Shock train - shear flow interaction}
\label{sec:2a}
  Figure.\ref{fig:shock_shear_interaction} shows a simple schematic of shock train flows in an isolator with a cavity. After interacting with the incoming boundary layer, a normal shock undergoes a bifurcation process at a freestream Mach number greater than 1.5. The location of the shock train is controlled by the ratio of upstream to downstream pressure (p$_b$/p$_{\infty}$) and duct blockage ratio ($\beta$). Each successive shock structure includes a short compression and re-acceleration region. As described earlier in Sec.\ref{sec:label1}, the shock train is subject to self-excited oscillation due to local pressure perturbations. At the same time, the supersonic flow across the cavity results in a self-sustained oscillation due to the interaction between the shear layer and acoustic disturbances from the cavity front wall. The combination of this low-frequency and high-frequency oscillatory flow may correlate, or one of the two flows may dominate the flow field. 

\subsection{Geometrical details}
\label{sec:2b}
\begin{figure}
  \includegraphics[width=0.7\linewidth]{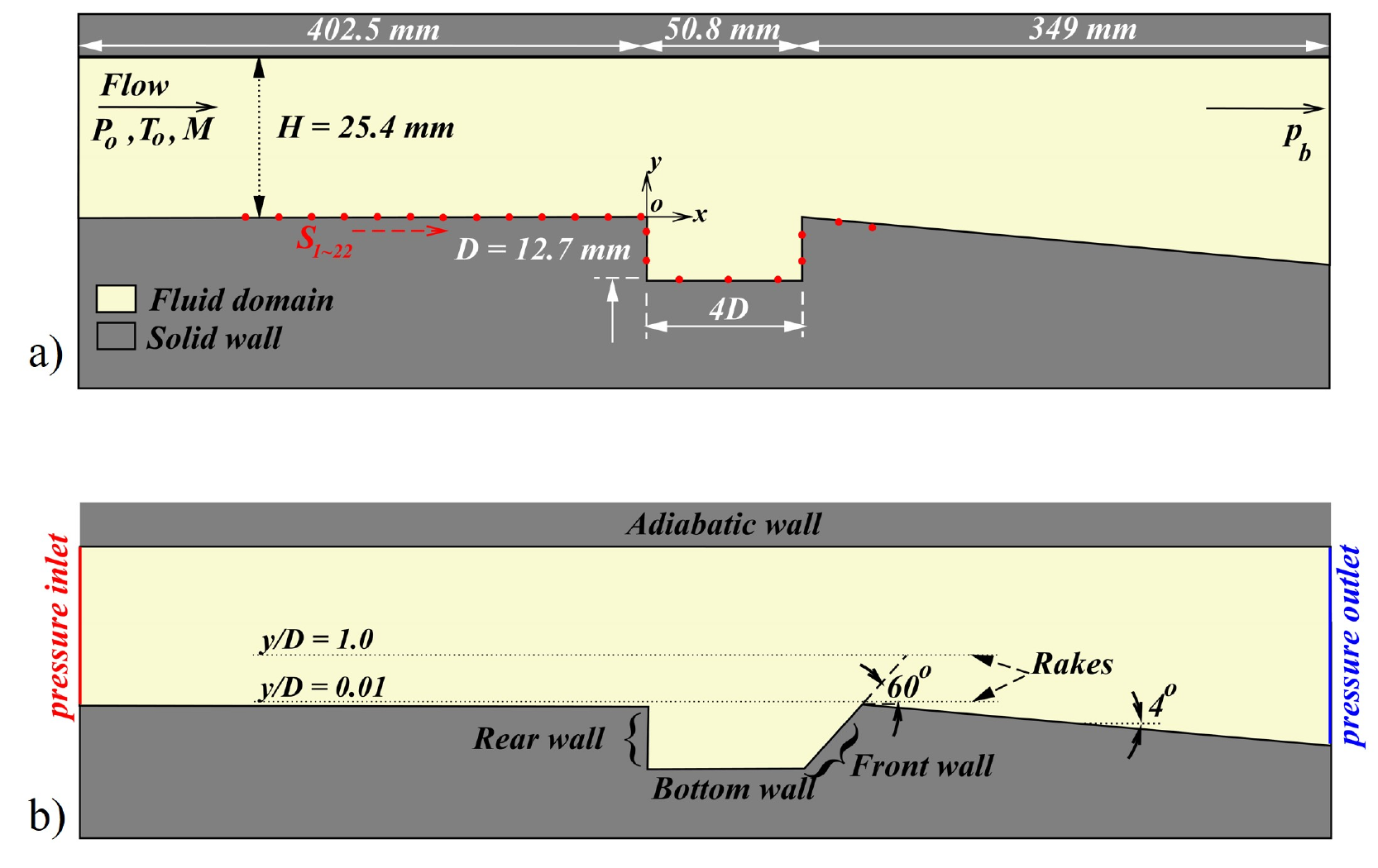}
  \caption{\label{fig:geometry} Schematic of the present numerical domain along with the geometrical details, boundary conditions, and pressure probing stations ($S_1 \sim S_{22}$).  (not to scale)}
  \end{figure}

  The geometrical details of the isolator section with cavity and downstream diffuser are taken from the experimental work of Micka et al.\cite{micka_2009} and Fotia et al.\cite{fotia_2013}. The isolator section is a constant area duct with a height (H) of 25.4 mm and a length (L$_{iso}$) of 402.5 mm. The cavity section is an open type with a length (L) to depth (D) ratio [L/D] of 4 and a depth (D) of 12.7 mm. The downstream portion of the cavity is a single-sided diffuser with a lower wall divergence angle of 4$^o$ and a divergence section length (L$_{div}$) of 349.0 mm. The origin of the present geometry is considered at the leading edge of the cavity. The geometrical details are shown in Fig.\ref{fig:geometry}. The measurements are taken at various locations. The locations where the wall pressure data were measured are labeled as `S' with a point number (1$\sim$22). In addition, the streamwise rakes ranging from -20.0 $<$ [x/D] $<$ 15.5 for [y/D] = 0.01 and 1.0 are extracted to understand the influence of the interaction between the shock train and the shear layer.
  
\subsection{Flow configuration}
\label{sec:2c}
  The present numerical simulations are performed at a freestream Mach number (M$_{\infty}$) of 2.2. The boundary layer is allowed to grow from the inlet region. The value of boundary layer thickness ($\delta$/D) is 0.24 at (x/D) = -20.0 from the bottom wall obtained after the steady-state solution is reached. The boundary layer thickness value is taken as the height at which the velocity reaches 98\% of the freestream velocity (u$_{\infty}$= 544 m/s). The back pressure ratio [p$_b$/p$_{\infty}$ = 0.7 (no shock train), 5.0, and 6.0] is changed to obtain different flow conditions in the isolator. Other important input parameters can be found in Table.\ref{tab:table1}.
 
  \begin{table}
  \caption{\label{tab:table1} Numerical input parameters.  }
  \begin{ruledtabular}
  \begin{tabular}{lc}
  Parameters & Values \\
  \hline
  Isolator length (\textit{L$_{iso}$, mm})                      &  402.5        \\    
  Isolator height (\textit{H, mm})                              &  25.4         \\
  Cavity length (\textit{L$_{cav}$, mm})                        &  50.8         \\
  Cavity depth (\textit{D, mm})                                 &  12.7         \\
  Boundary layer thickness (\textit{$\delta$/D})                &  0.24         \\
  Inlet total pressure (\textit{P$_{01}$, kPa})                 &  303          \\
  Outlet static pressure (\textit{p$_{b}$, kPa})                &  20, 142, 170 \\
  Inlet total temperature (\textit{T$_{01}$, K})                &  300          \\
  Mach number (\textit{M$_\infty$}, at x/D = -31.7)             &  2.2          \\
  Inlet Reynolds number (\textit{Re} x \textit{10$^{-7}$, /m})  &  2.0          \\
  \end{tabular}
  \end{ruledtabular}
  \end{table}
\section{NUMERICAL METHODOLOGY}
\label{sec:3}
  The interaction of the shock train with the cavity shear layer is analyzed numerically using a commercial fluid dynamics package of ANSYS Fluent\textsuperscript \textregistered. The unsteady statistics of the shock train flow field and the cavity shear layer flow are accurately estimated using higher-order numerical techniques by direct numerical simulation (DNS) and large eddy simulation (LES). LES requires a rigorous grid configuration near the wall to resolve the boundary layer, which is computationally expensive. Alternatively, a hybrid Reynold's Averaged Navier-Stokes (RANS)/LES-based method called Detached Eddy Simulation (DES), can resolve turbulent features in the largely separated flow with low computational cost and greater flexibility in grid generation \cite{spalart_2000}. DES provides a RANS-based solution within the boundary layer region and switches to LES in the separated boundary layer flows \cite{spalart_2006, shur_2008}. Researchers performing simulations of cavity flow using a hybrid RANS/LES method were extensively reviewed by Lawson et al. \cite{lawson_2011}. They concluded that DES simulation provides a better comparison with experimental results. Therefore, in the present study, a similar numerical methodology is applied to determine the unsteady dynamics of the shock train and the cavity shear layer interaction in a scramjet isolator. Moreover, the shock train flows are mainly affected by the streamwise component or the longitudinal component along the symmetry plane \cite{ram_2021b}. Recent numerical studies on two-dimensional cavities have provided sufficient insight into the shear layer flows in the cavity compared to the experimental results \cite{aravind_2019, karthick_2021, aravind_2021}. Therefore, the present numerical study focused on the two-dimensional flow field, while the effects in the span (i.e., lateral instability, flow separation in the corner, and sidewall effects) were not considered.
  
\subsection{Numerical procedure and boundary conditions}
\label{sec:3a}
  A two-dimensional planar domain is computed along with the cavity, as described in Sec.\ref{sec:2b}. The initial time-averaged solution is obtained with a RANS-based simulation, and the near-wall turbulence is resolved by the k-$\omega$ SST model. The spatial and temporal discretization is performed using a second-order upwind scheme and a fully implicit density-based method. The convective fluxes across each grid cell are evaluated using the Advection Upstream Splitting Method (AUSM) together with the Total Variation Diminishing (TVD) slope limiter, which uses the minmod function to limit the overshoots and undershoots at the cell faces \cite{fluent_2010}. After a steady-state solution was obtained, the solver proceeded to the transient simulation. To improve accuracy, the transient formulation is chosen as the second-order implicit scheme. The additional advantage of solving the time integration terms with the implicit scheme is that it is unconditionally stable with respect to the time step size. Since the unsteady simulation is initialized from a steady-state solution, a saturation time is required to develop a transient flow in the computational domain and to obtain self-oscillating shock train flows in a constant area duct \cite{ram_2022}. A detailed discussion of the saturation time can be found in Sec.\ref{sec:3c}.  

  The working fluid is assumed to be perfect air obeying the ideal gas law. The viscosity of the fluid is a function of temperature, which is determined using Sutherland's law with the three-coefficient method. The transient simulation is performed using the SST k-$\omega$ based DES model with the standard constants given in the solver. The model equations are as follows
  
  \begin{equation}
   \label{eq:k_eqn}
   \frac{\partial (\rho k)}{\partial t}+ \frac{\partial(\rho k u_i)}{\partial x_i} = 
   \frac{\partial}{\partial x_j} \Bigg[\Bigg(\mu + \frac{\mu_t}{\sigma_k}\Bigg)\frac{\partial k}{\partial x_j} \Bigg] + G_k - \Delta_k
  \end{equation}

  \begin{equation}
   \label{eq:w_eqn}
   \frac{\partial (\rho \omega)}{\partial t}+ \frac{\partial(\rho \omega u_i)}{\partial x_i} = 
   \frac{\partial}{\partial x_j} \Bigg[\Bigg(\mu + \frac{\mu_t}{\sigma_\omega}\Bigg)\frac{\partial \omega}{\partial x_j} \Bigg] + G_\omega - \Delta_\omega +D_\omega
  \end{equation}

  where G$_k$ represents the generation of turbulence kinetic energy due to mean velocity gradients.  The G${_\omega}$ accounts for the production of $\omega$. ${\Delta_k}$ and ${\Delta_\omega}$ represent the dissipation of k and $\omega$ due to turbulence, respectively. ${\sigma_k}$ and ${\sigma_\omega}$ are the turbulent Prandtl numbers for k and $\omega$, respectively. For DES simulation, the dissipation of kinetic energy (${\Delta_k}$) is modified. Such that
  
  \begin{equation}
   \label{eq:diss_k}
   \Delta_k = \rho \beta^* k \omega F_{DES}
  \end{equation}
  where F$_{DES}$ is the maximum of ratio of turbulent length scale (L$_{t}$) to the local grid spacing ($\Delta x$, $\Delta y$) and 1.0. The turbulent length scale is defined from the RANS model and is given as 
  \begin{equation}
   \label{eq:des_L}
   L_t = \frac{\sqrt k}{\beta^* \omega}
  \end{equation}

  Together with the above model equation, it is necessary to introduce the solution with realistic turbulence parameters. The turbulence intensity at the inlet and the length scale are given as 2\% and 0.0254 m (i.e., duct height, H) respectively. To achieve smooth convergence and stability, the CFL number and under relaxation factor are set to a lower value and linearly increased with iteration for steady-state simulation using a full multi-grid (FMG) initialization method. The residuals of the conservation equation and turbulence equation were brought down below 10$^{-6}$ for steady-state simulations. After reaching the steady-state solution, a transient simulation is performed. The time step size of 1.0e$^{-6}$ s was chosen based on the physical convection time of an air molecule moving along the duct. The physical time ($\Delta_{pt}$) can be expressed as the ratio between the length of the isolator (L$_{iso}$) and the freestream velocity (u$_{\infty}$), which is about 7.54e$^{-4}$ s (based on L$_{iso}$ = 0.4025 m and u$_{\infty}$ = 544.0 m/s). The chosen time step size is much smaller than the physical time to ensure the prediction of unsteady dynamics. The inlet region of the isolator is specified as the pressure inlet, for which the stagnation properties P$_{01}$ and T$_{01}$ are specified. The outlet boundary is considered as a non-reflecting pressure outlet boundary condition to specify the static pressure or back pressure (p$_b$). The isolator and cavity walls are considered as adiabatic walls with no-slip boundary conditions.
  
\subsection{Grid independence study}
\label{sec:3b}
   \begin{figure}[b]
  \includegraphics[width=0.5\linewidth]{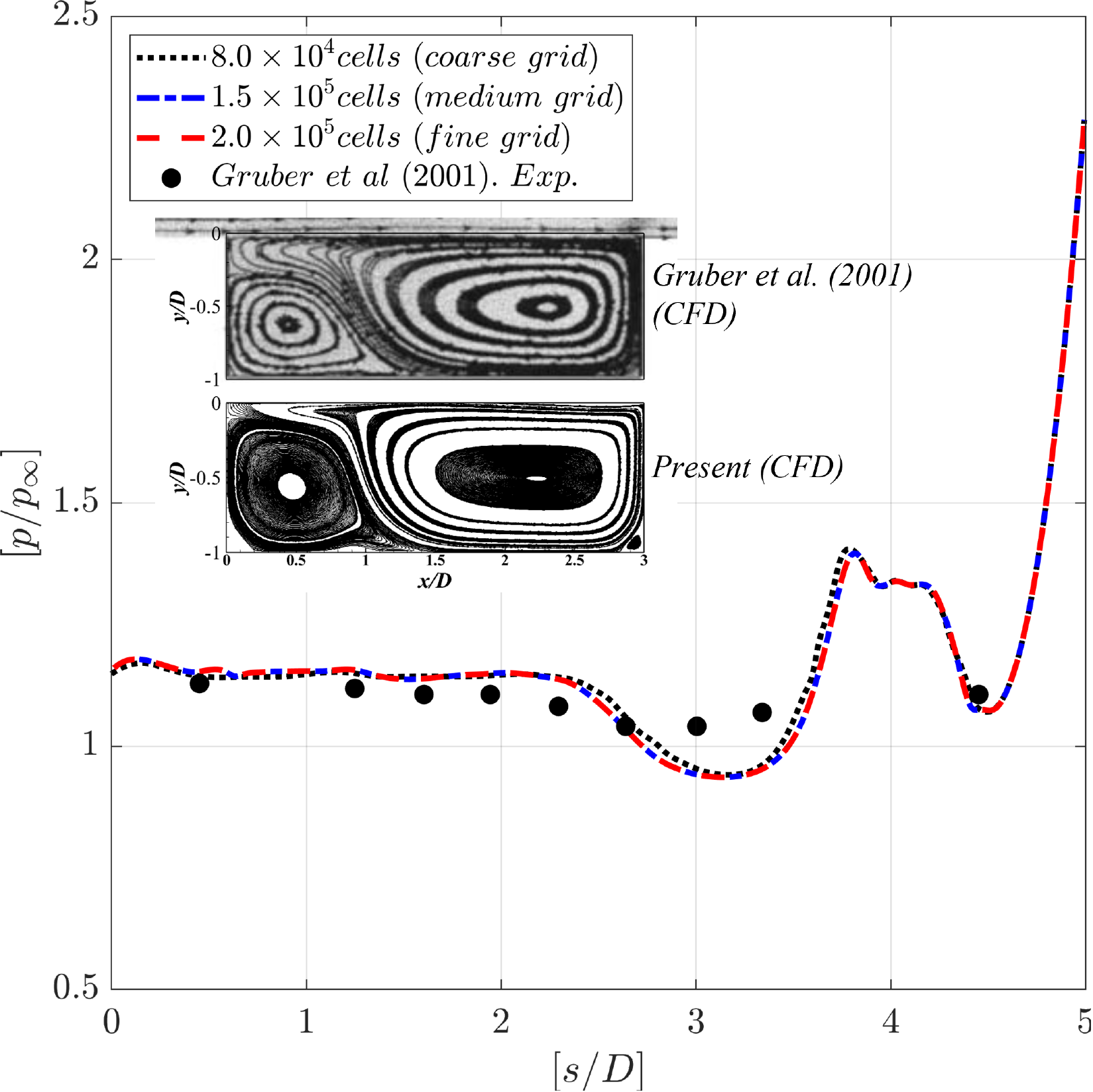}
  \caption{\label{fig:validation_gruber} Comparison of wall static pressure (\textit{p/P$_{\infty}$}) along the cavity wall surface (\textit{x/H}) with the experimental pressure distribution of Gruber et al.\cite{gruber_2001} for the following freestream values: \textit{P$_{0}$} = 690 kPa, \textit{T$_{0}$} = 300 K, \textit{M$_{\infty}$} = 3.0. Inside figure: The streamline inside the cavity shows two large counter-rotating vortices in comparison with Gruber et al.\cite{gruber_2001}}
  \end{figure}

 \begin{table*}
  \caption{\label{table:grids} Grid counts}
\begin{ruledtabular}
 \begin{tabular}{l c c c c c c}
  Grid & Domain	& Cavity &  First cell & Cell growth & Toal no. of \\
   & N$_x$$\times$N$_y$ & N$_x$$\times$N$_y$ & distance ($\Delta$/D) & ratio & cells \\
  \hline
  coarse   &  400$\times$150  &  200$\times$100   & 4.0 x 10$^{-4}$ &  1.35 &   8.0 x 10$^4$  \\
  medium   &  600$\times$175  &  250$\times$140   & 2.4 x 10$^{-4}$ &  1.25 &   1.5 x 10$^5$  \\
  fine     &  800$\times$200  &  300$\times$160	& 8.0 x 10$^{-5}$ &  1.20 &   2.0 x 10$^5$  \\
  present  &  1500$\times$200 &  300$\times$150	& 8.0 x 10$^{-5}$ &  1.20 &   3.5 x 10$^5$  \\
  \end{tabular}
  \end{ruledtabular}
  \end{table*}   

\begin{figure}
  \includegraphics[width=0.5\linewidth]{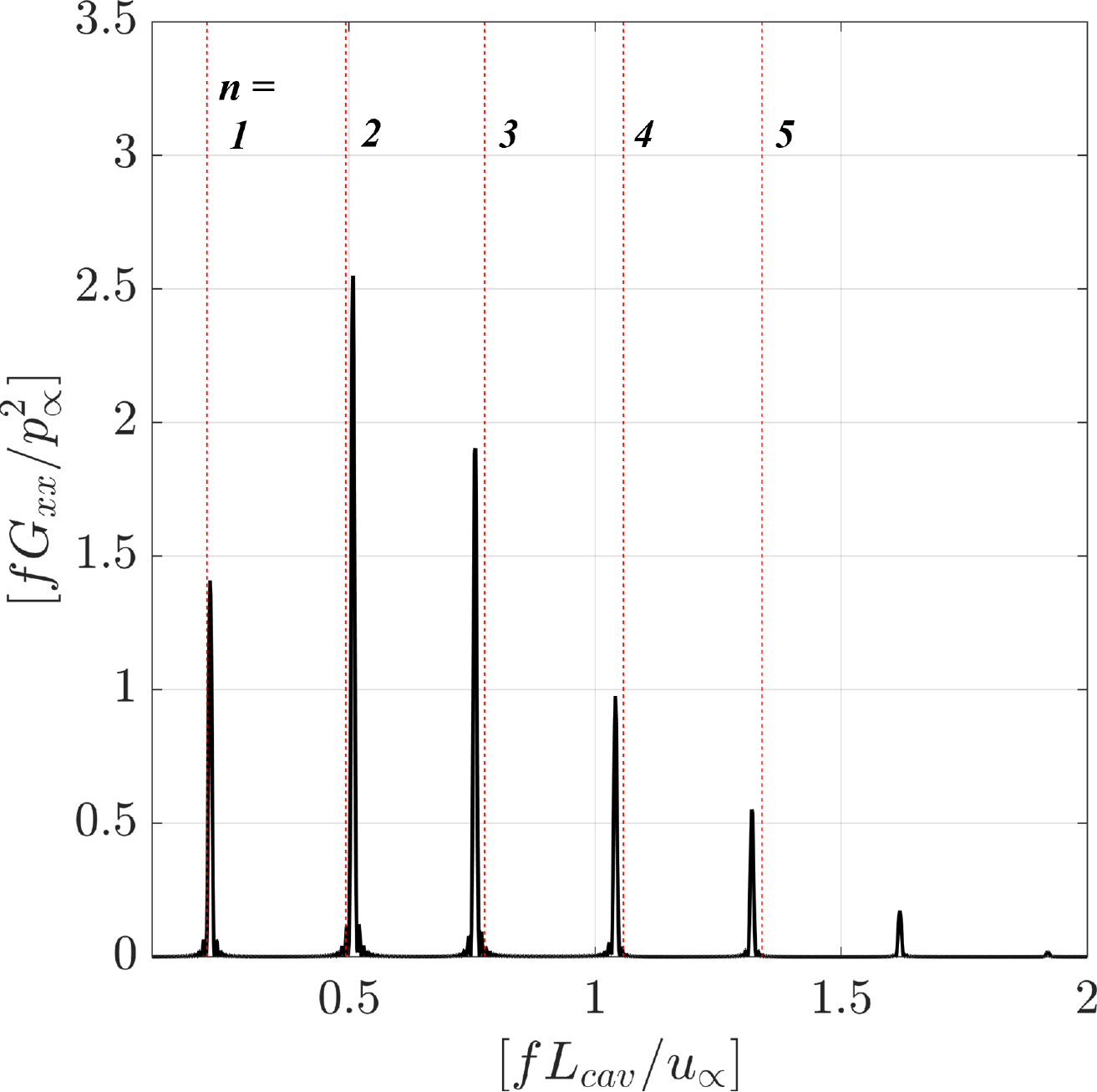}
  \caption{\label{fig:gruber_psd} Power spectrum obtained from wall static pressure (p/p$_{\infty}$) at [x/D] = 1.5 from fine grid. n is the mode number(1$\sim$5). red dashed line represents the modes obtained from the modified Rossiter's formula (Eq.\ref{eq:ross}). }
  \end{figure}

  For the purpose of a grid independence study, a two-dimensional planar open cavity is adopted from Gruber et al [59]. The cavity depth [D] is 8.9 mm and the length-to-depth ratio (L$_{cav}$/D) is 3.0. The grid is created using Pointwise commercial grid generation software and the near wall region is clustered to resolve the boundary layer flow. The accuracy of the numerical solution depends strongly on the distribution of grid cells in the fluid domain. The maximum equiangular skewness of the grid cells is kept below 0.3. To ensure that the present numerical methodology is independent of the grid resolution, three different grid levels were generated and tested. The details of the grid number can be found in Table.\ref{table:grids}. The total number of cells in the coarse, medium and fine grids are 8.0E4, 1.5E5, and 2.0E5, respectively. The cavity wall static pressure is plotted along the surface length and compared with the experimental pressure distribution (see Fig.\ref{fig:validation_gruber}). The static pressure distribution between the three grid levels does not show any difference, and the numerical solution shows clear grid independence with medium to fine grid levels. The deviation of the pressure distribution in comparison with experimental values is less than 5\% and the differences are mainly due to the three-dimensionality or the uncertainties in the experiments \cite{karthick_2021}. Therefore, a grid similar to the fine grid level is sufficient to capture the flow field and is adopted for the present numerical cases, and the grid spatial counts along the x and y directions are given in Table.\ref{table:grids}. In order to capture the shock train flow and cavity shear layer, extreme care was taken in the grid generation. For the unsteady simulation, a time step size of ($\Delta t/T$ = 1e$^{-3}$) is chosen and simulated for the above cavity case. The static wall pressure at [x/D] = 1.5 (corresponding to [s/D] = 2.5) is monitored to obtain the dominant frequency of the cavity oscillation. The frequency spectra of the self-excited oscillation can be estimated using the semi-empirical formula of Rossiter \cite{rossiter_1964} and Heller et al \cite{heller_1996}. It has been modified to account for the effect of a higher Mach number on the oscillation frequency. Fig.\ref{fig:gruber_psd} shows the dimensionless power spectrum ($fG_{xx}$(p)/(p$_\infty$)$^2$) compared with the modified Rossiter formula defined by equation (Eq.\ref{eq:ross}).
   
   \begin{equation}
   \label{eq:ross}
   \frac{f L_{cav}}{u_{\infty}} = \frac{n-\alpha}{M_{\infty}(1+[(\gamma-1)/2]M_{\infty}^2)^{-1/2}+1/k}
  \end{equation}
  where \textit{f} is the frequency in Hz, L$_{cav}$ is the length of the cavity, n is the mode number for the cavity oscillation from the peak frequencies, 'k' is an empirical constant defined by the convection velocity versus the free stream velocity, and $\alpha$ is the phase delay. The constants k and $\alpha$ for the validation case are 0.57 and 0.25 respectively adopted from Rossiter et al.\cite{rossiter_1964}. The values of the calculated peak frequency and that calculated from Rossiter's formula are tabulated in Table.\ref{tab:ross_modes}, which shows that the difference is less than 5\% for the first five modes.

  \begin{table}
  \caption{\label{tab:ross_modes} Comparison of cavity oscillation frequency with Rossiter's empirical formula (Eq.\ref{eq:ross})}
  \begin{ruledtabular}
  \begin{tabular}{l c c }
  mode number 	& Present CFD mode        & Rossiter's mode          \\
  (n)             & \textit{fL$_{cav}$/u$_{\infty}$} & \textit{fL$_{cav}$/u$_{\infty}$}  \\
  \hline
  1  &  0.2172  &  0.2114       \\
  2  &  0.5874  &  0.4933       \\
  3  &  0.7567  &  0.7753  	  \\
  4  &  1.0412  &  1.0572  	  \\
  5  &  1.3180  &  1.3391  	  \\
  \end{tabular}
  \end{ruledtabular}
  \end{table}
  
\subsection{Unsteady numerical error estimation}
\label{sec:3c}
  \begin{figure*}
  \includegraphics[width=1\linewidth]{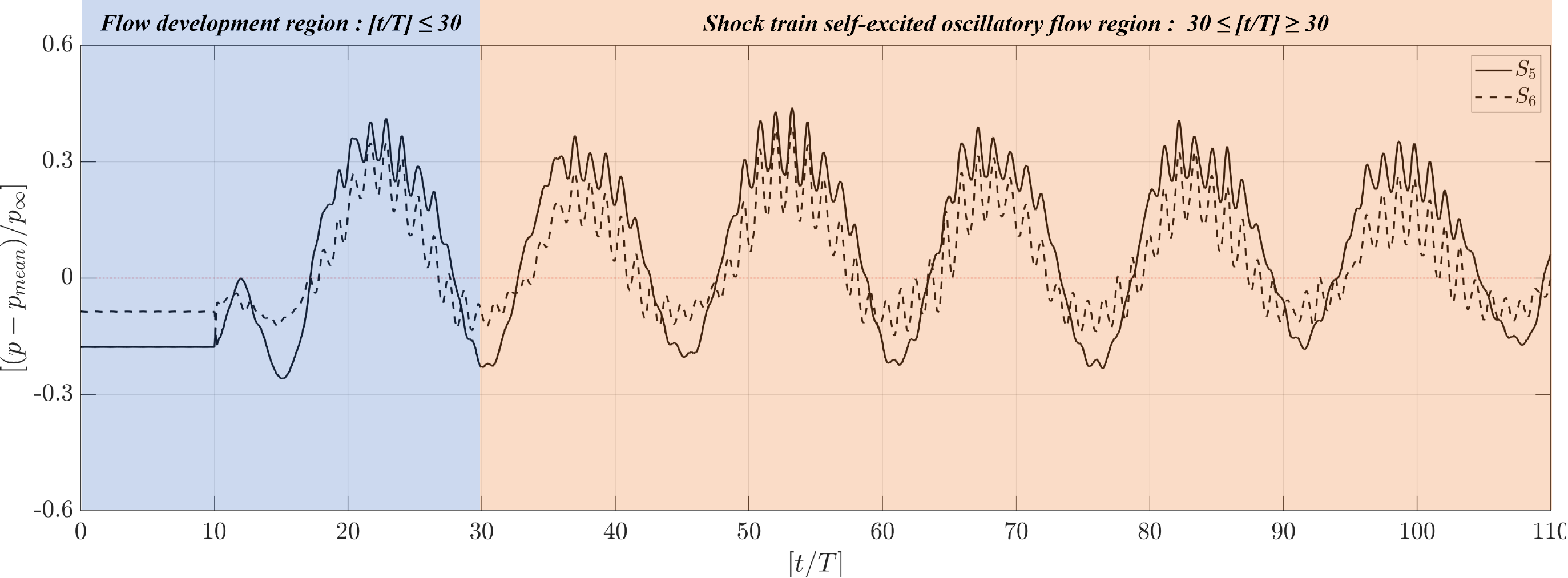}
  \caption{\label{fig:probe_wall_press} Probed wall static pressure [(p-p$_{mean}$)/p$_{\infty}$] measurements along the bottom wall surface (y/H= 1.0) at S$_5$ and S$_6$ (i.e., x/D = 18.89 and 20.5) for the case (p$_b$/p$_{\infty}$=6.0). The flow development time is required during a numerical run switching from steady to unsteady simulation. Self-excited oscillatory flow is observed after the time [t/T] $\geq$ 20. Data sampling for unsteady statistics is carried out after [t/T] $\geq$ 30. The maximum to minimum pressure extent shows the shock train oscillatory motion.}
  \end{figure*}
  Large-scale unsteady simulation is bound to accumulation of error due to numerical discretization and can be controlled by reducing the length scale to an infinitely small value. Smirnov et al.\cite{smirnov_2015} proposed a comprehensive methodology to sort out the grid size effect on numerical precision and error estimation. The poor grid resolution and time step size could lead to the accumulation of numerical error within the unsteady simulation. The detailed method for the estimation of numerical reliability can be found in Refs. \cite{smirnov_2015}. This method is well suited for the application of scramjet simulations and is adopted by various researchers recently \cite{su_2018,jiao_2018,ram_2021b}. Hence in the present work, the author evaluates the total relative error of integration by assuming maximum allowable error limits of 5\%. The reliability (R$_s$) is the ratio of the maximum allowable number of time steps (n$_{max}$) to the number of simulated time steps (n$_{sim}$). From the present grid resolution, the maximum allowable number of time steps (n$_{max}$) is 1 $\times$ 10$^{11}$. To resolve the low-frequency oscillation of the shock train, a physical time of up to 50ms is simulated. Thus, the reliability (R$_s$) of the present cases is 1 $\times$ 10$^{6}$, which clearly shows that the present simulation has high reliability on unsteady simulation. The unsteady wall static pressure measurement from the present numerical simulation probed at two different points (S$_5$ and S$_6$) is plotted and shown in Fig.\ref{fig:probe_wall_press}. In the present numerical calculation, a stable unsteady shock oscillatory flow is obtained for time, [t/T] $\geq$ 20. Therefore, to resolve the unsteady dynamics, the data obtained about 50 ms after the saturation time ([t/T] $\geq$ 30) are processed. 

\subsection{Post Processing}
\label{sec:3d}
  Post-processing is performed using MATLAB R2021b\textsuperscript\textregistered. The Fast Fourier Transform (FFT), two-point cross-correlation, and Dynamic Mode Decomposition (DMD) analysis are performed to extract the transient dynamics. From the two-dimensional DES simulation, the set of data for each node at every time step is exported in ASCII format. Due to the large size of the data, the exported scattered data is converted into uniform gridded data with little or no loss of spatial information. The spatial data are extracted only for a specific region that includes the shock train and the cavity (-20 $\leq$ [x/D] $\leq$ 15 and -1 $\leq$ [y/D] $\leq$ 2) and interpolated to a coarse grid of points of equal distance (1001 $\times$ 101) that is used for modal analysis. On the other hand, a set of scattered point data (S$_1$ $\sim$ S$_{22}$) and two probing rakes ([y/D] = 0.01, 1.0) are extracted and used for FFT and cross-correlation analysis.
  
\section{RESULTS AND DISCUSSIONS}
\label{sec:4}
  Numerical analysis was carried out to study the interaction between the shock train and the cavity shear layer in a scramjet isolator. Cases were considered with three different pressure ratios from upstream to downstream, controlled by changing the outlet pressure or the back pressure (see Table.\ref{tab:cases} ), and two different cavity geometries (see Fig.\ref{fig:geometry}). The cases can be distinguished into the case without shock train, shock train ends before the cavity leading edge and shock train near the cavity location. Even with a fixed pressure ratio, the shock train experiences a self-excited oscillation and interacts with the cavity shear layer flow. These flow features are discussed in the following sections.
  
\begin{table}
  \caption{\label{tab:cases} Case details}
  \begin{ruledtabular}
  \begin{tabular}{l c c c}
  Cases & Back Pressure	& (p$_b$)/(p$_{\infty}$) & Cavity front wall  \\
   & (p$_b$), kPa &  & angle [$\theta$], deg \\
  \hline
  case 1   &  20   &  0.7  & 90   \\
  case 1a  &  142  &  5.0  & 90   \\
  case 1b  &  170  &  6.0	 & 90   \\
  case 2   &  20   &  0.7	 & 60   \\
  case 2a  &  142  &  5.0	 & 60   \\
  case 2b  &  170  &  6.0	 & 60   \\
  \end{tabular}
  \end{ruledtabular}
  \end{table}
  
\subsection{Steady and Unsteady statistics}
\label{sec:4a}
  \begin{figure}
    \subfigure[]{\label{fig:pres_dist_a}\includegraphics[width=0.45\linewidth]{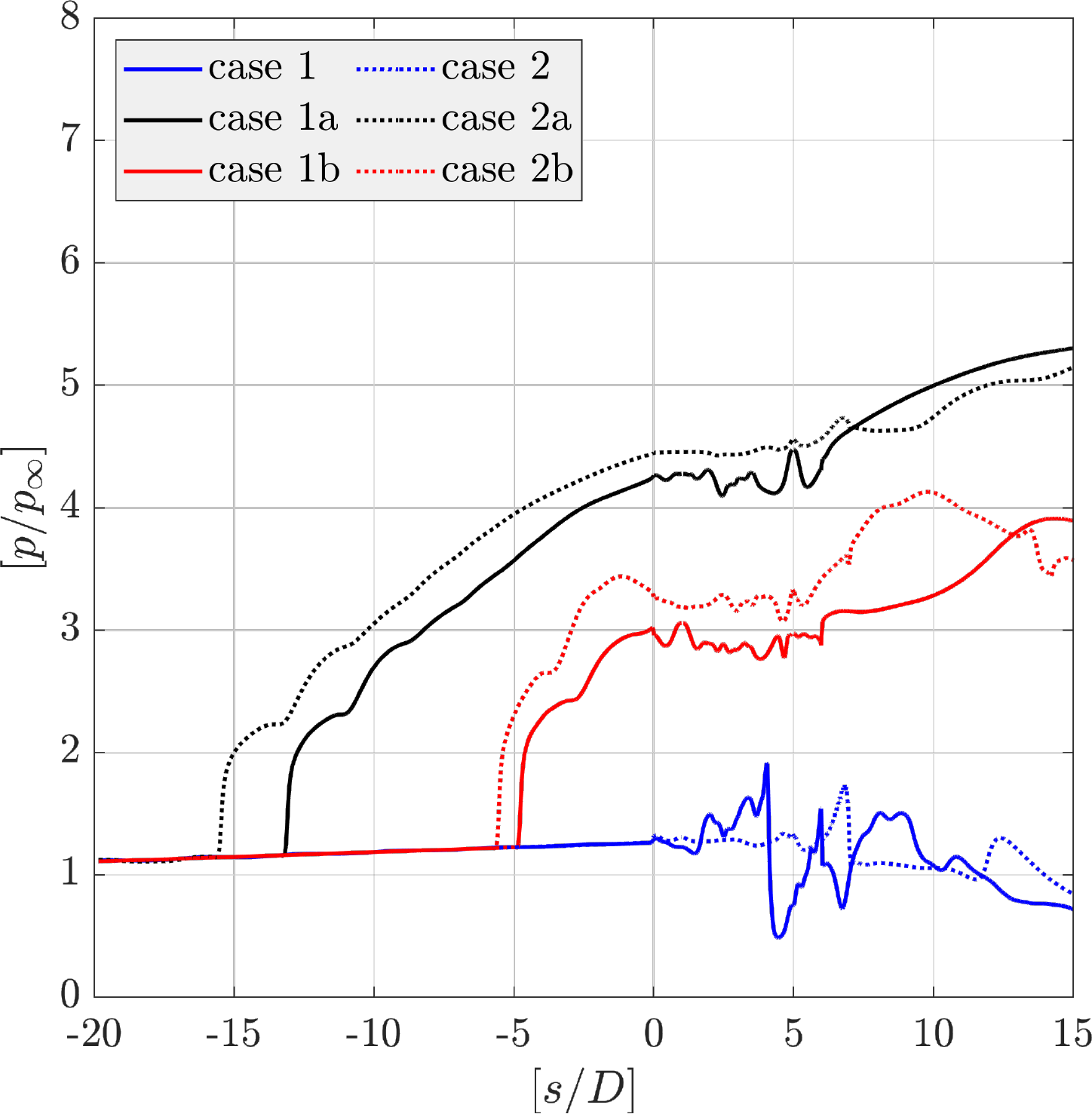}}
    \subfigure[]{\label{fig:pres_dist_b}\includegraphics[width=0.45\linewidth]{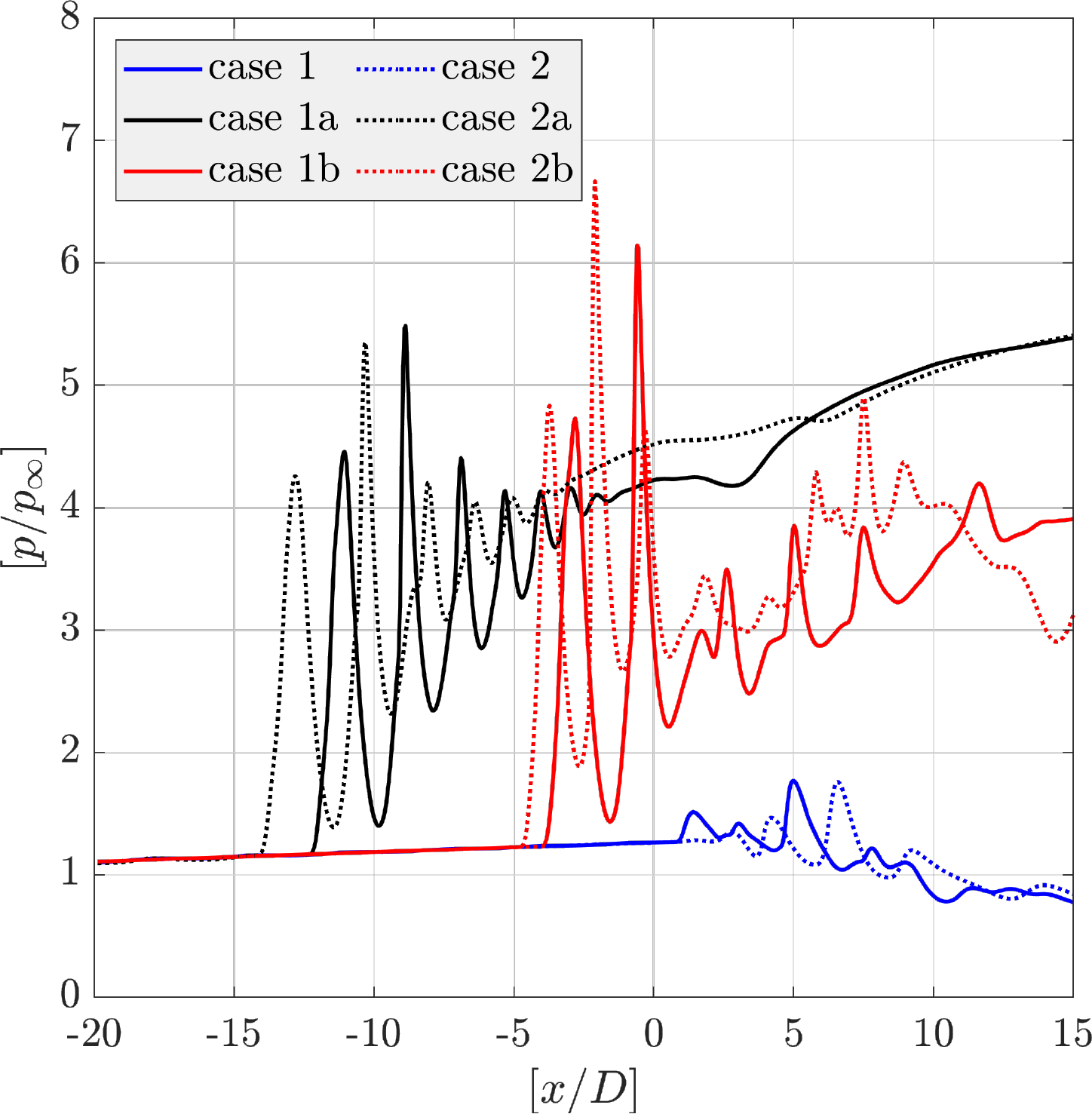}}
    \caption{\label{fig:pressure_distribution} Time-averaged static pressure measurements from RANS simulation are plotted along (a) bottom wall [y/D] = 0, (b) centerline [y/D] = 1.0. The solid line represents 90$^o$ }
  \end{figure}

  The time-averaged static pressure along the centerline (y/D = 1.0) and the static pressure distribution at the bottom wall extracted from the steady-state RANS k-$\omega$ SST simulation for different cases are compared and shown in Fig.\ref{fig:pressure_distribution}. The static pressure distribution for the cavity front wall angle of 90$^o$ is shown as a solid line and the case of 60$^o$ is shown as a dotted line. The wall static pressure comparison among various cases is shown in Fig.\ref{fig:pres_dist_a} where the x-axis is defined as the surface length, and the origin (s/D = 0) refers to the cavity leading edge point. The pressure comparison graph of cases (1a \& 1b) shows that the location of the shock train is slightly lower than the case (2a \& 2b). This is because the angle of the cavity front wall of 60$^o$ makes the exit area smaller than the 90$^o$ case. So, for the given back pressure (p$_b$), the shock train leading shock moves to a forward location to match the given back pressure at the outlet. From Fig.\ref{fig:pres_dist_b}, it can be seen that the shock train in cases (1a \& 2a) ends before the cavity leading edge. This corresponds to a supersonic inlet operating at ramjet mode, where the combustion process occurs at subsonic velocity. In cases (1b \& 2b), the shock train located above the cavity shows strong interaction with the cavity shear layer and observed distorted shock peaks. The back pressure ratio (p$_b$/p$_{\infty}$ =0.7) shows supersonic flow around the cavity in a confined flow field without shock interactions. The centerline pressure distribution shows a weak oblique shock originating from the front wall of the cavity.

  \begin{figure}
    \subfigure[]{\label{fig:mach_instant}\includegraphics[width=0.45\linewidth]{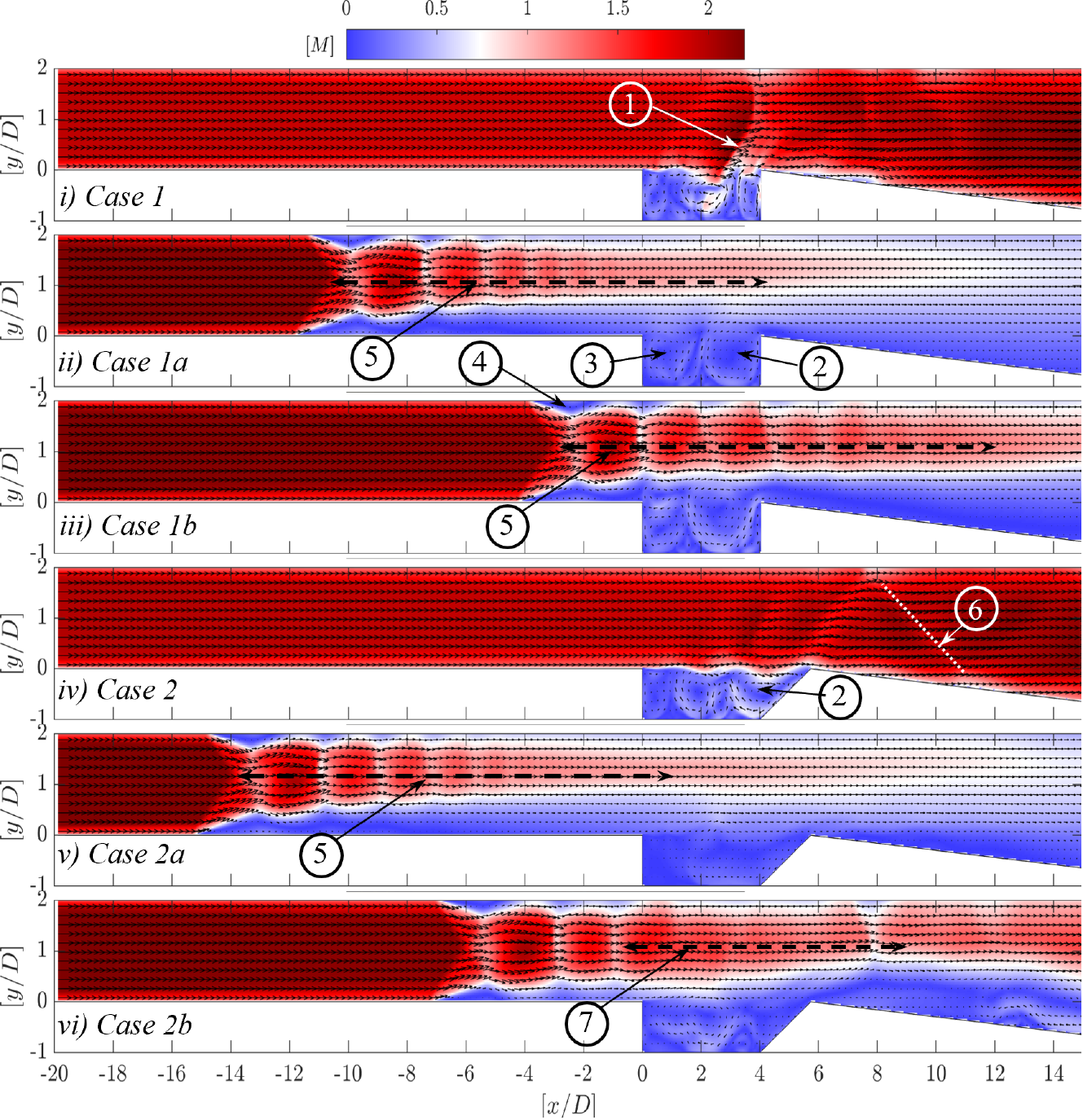}}
    \subfigure[]{\label{fig:vort_instant}\includegraphics[width=0.45\linewidth]{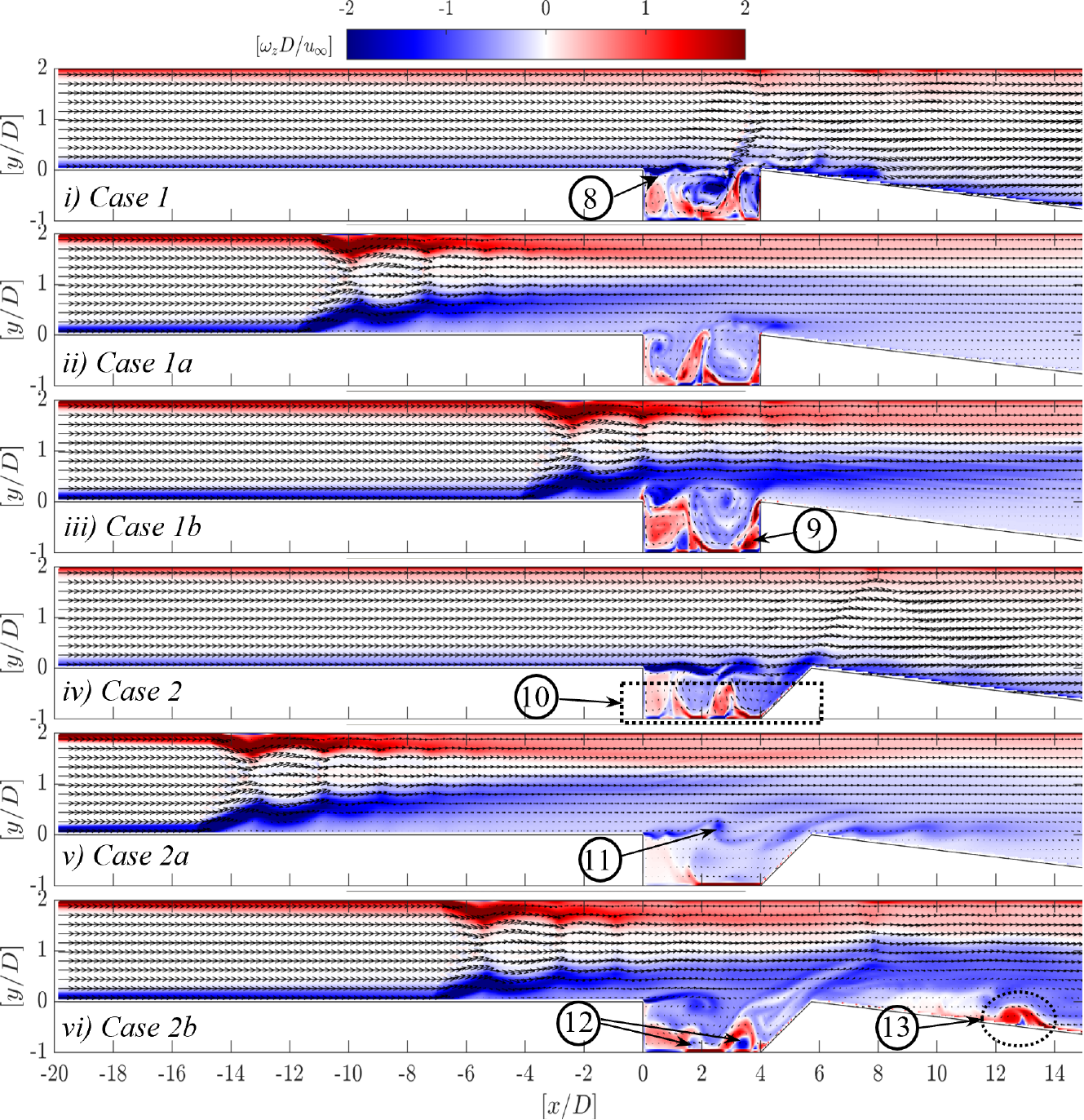}}
    \caption{\label{fig:mach_vorticity} Instantaneous flow field showing (a) Mach number [M] and (b) normalized vorticity [$\omega D/u_{\infty}$] contour. (\textit{i$\sim$vi}) represents case 1,1a,1b,2,2a,2b respectively and details shown in Table.\ref{tab:cases}. Various distinct flow features marked are (1) shear layer induced weak shock wave, (2) cavity primary recirculation, (3) cavity secondary recirculation (4) separated boundary layer due to shock train leading shock wave, (5) extent of shock train and mixing layer (pseudo shock wave), (6) reflected oblique shock wave, (7) smeared shock train due to shear layer interaction, (8) large scale vortical structures from shear layer, (9) small scale corner vortices, (10) interaction of counter-rotating vortices inside the cavity, (11) distorted large scale vortical structures travels downstream (12) small scale vortical structures in cavity recirculation region, (13) vortical structures from the downstream separated boundary layer}
  \end{figure}  

  Later, with the steady convergence results from the RANS simulation, DES combined with the k-$\omega$ SST formulation is switched on. The delayed detached eddy simulation (DDES) is chosen as the shielding function \cite{spalart_2006}. After switching to the unsteady simulation, a saturation time of [t/T] = 30 is run to achieve a stable unsteady solution, as previously described in Sec.\ref{sec:2c}. In order to understand the different flow features along the isolator with cavity flow, the instantaneous Mach contour and vorticity contour are shown in Fig.\ref{fig:mach_vorticity} along with the quiver representing the velocity vector. Fig.\ref{fig:mach_instant} and \ref{fig:vort_instant} (i, ii, iii) show the Mach contour [M] and normalized vorticity [$\omega D/u_{\infty}$] for the case where the cavity front wall has an angle of 90$^o$, and (iv, v, vi) for the case of 60$^o$. The main flow features such as the shear-induced shock wave, the shock train, and the recirculation region in the cavity can be clearly seen. The leading shock structure of the shock train is more oblique rather than normal along the core region of the flow. This is evident from the centerline pressure plot (see Fig. \ref{fig:pres_dist_b}), in which the pressure rise due to the first shock is less than that of the second shock wave. The formation of a normal shock train takes place when the freestream Mach number (M) $\geq$ 1.5, an oblique shock train (M $\geq$ 2.5), and at Mach number (M) between 1.7 and 2.2 represents a transition from a normal shock wave to an 'X' type shock train \cite{hunt_2018}. The transition from normal to oblique shock train may also depend on the boundary layer blockage ratio ($\beta$). For the present case, the boundary layer blockage ratio is calculated from the following equation (Eq.\ref{eq:blockage}),
  
  \begin{equation}
   \label{eq:blockage}
   \beta = 1- \Bigg[1- \frac{2 \delta ^*}{H}\Bigg]^2
  \end{equation} 

  where $\delta^*$ is the thickness of the boundary layer displacement measured at [x/D] = -20, and H is the height of the duct, giving the blockage ratio $\beta$ = 0.15. Controlling the blockage ratio by partially removing the boundary layer can form a normal shock train \cite{ram_2022}. It is also observed that the shock train near the cavity location undergoes smeared successive shock structure with the interaction of the cavity shear layer (see Fig.\ref{fig:mach_instant}v). This is due to the longitudinal driving force of the cavity shear layer. The Mach and vorticity contour also reveal the presence of a large recirculation region in the cavity with multiple counter-rotating vortices. Distinct distortions of the large-scale vortical structures are visible and marked as 11 in Fig.\ref{fig:vort_instant}v.
 
\subsection{Shock train and shear layer oscillatory characteristics}
\label{sec:4b} 

  \begin{figure}[hb]
\includegraphics[width=1\linewidth]{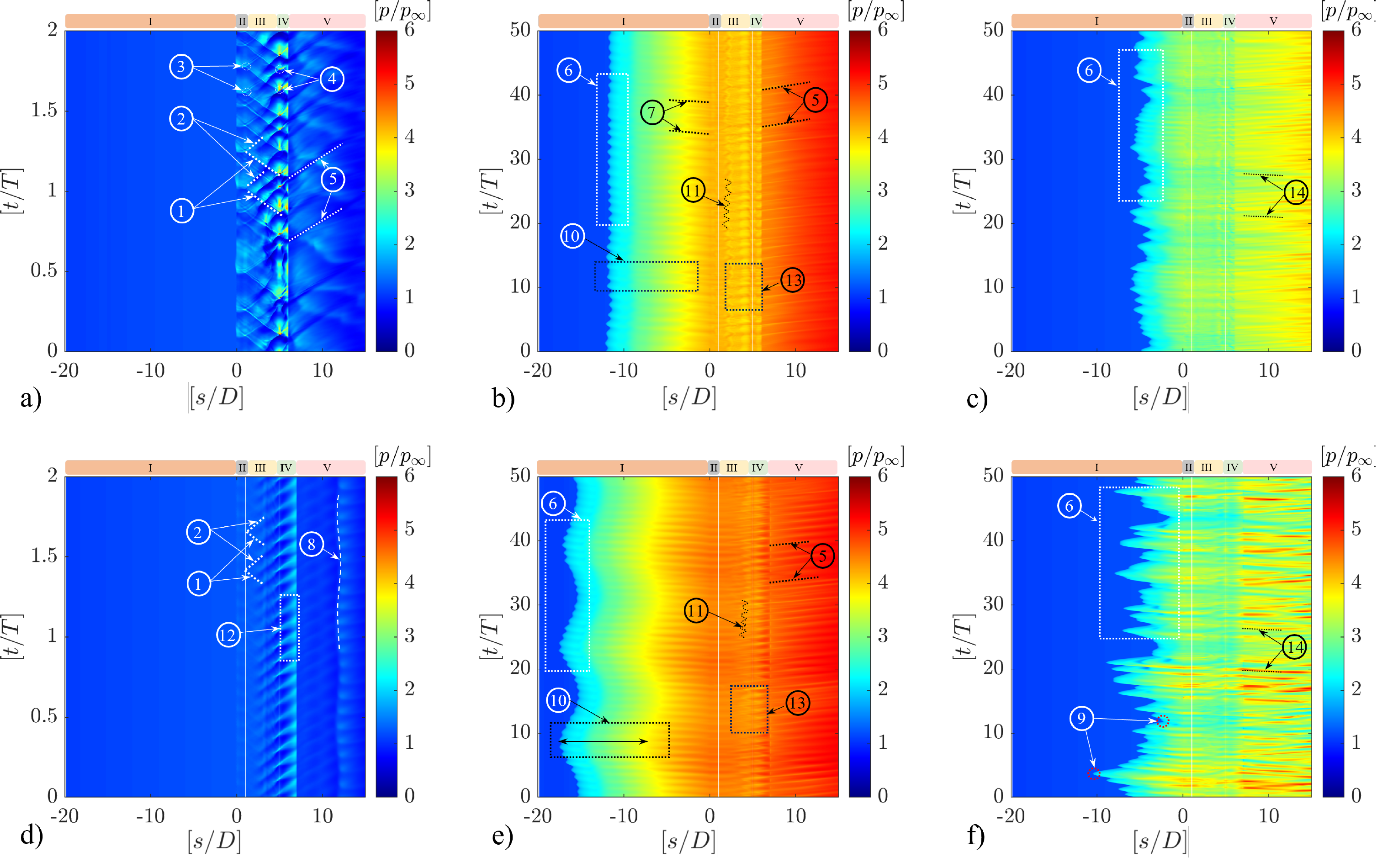}
  \caption{\label{fig:wall_press_xt} Non-dimensional static pressure [p/p$_{\infty}$] trace over time [t/T] along the bottom wall surface length [s/D].    [(a)-(c)] \& [(d)-(f)] shows cavity with front wall angle is 90$^o$ and 60$^o$ respectively. I) isolator section, II) cavity rear wall, III) cavity bottom wall, IV) cavity front wall and V) diffuser section. Observed flow features are 1) left running wave starts from front wall corner and ends at rear wall corner point, 2) reflected wave from the rear wall corner point, 3) \& 4) flow converging point of right running and left running wave at rear wall and front wall corner point respectively 5) disturbance propagating downstream in the diffuser section 6) trace of shock train oscillation 7) disturbance propagation towards isolator section 8) trace of separated flow in the diffuser section 9) upstream and downstream location of shock train, 10) maximum extent of shock train from wall pressure, 11) coalescence of right and left running wave, 12) \& 13) pressure rise due to shear layer impingement on front wall surface, 14) upstream propagation of disturbance from separated flow in the diffuser section.}
  \end{figure}

  The transient features extracted from the two-dimensional DES solution for a period of 30 $\leq$ [t/T] $\leq$ 80 (T = 1ms), correspond to a physical time of 50ms. The extracted data from the centerline rake and several probing stations (S$_1$ $\sim$ S$_{22}$) are arranged as a three-dimensional matrix for post-processing, as mentioned in Sec.\ref{sec:3d}. The space-time contour (i.e., x-t) of the static pressure distribution on the wall can be plotted to visualize the shock train and oscillatory motion in the cavity. This can also help to understand the mechanism of propagation of the pressure disturbance. The oscillation of the shock train is caused by local pressure fluctuations in the boundary layer flow. Any small disturbance that occurs in the downstream region propagates upstream through the subsonic boundary layer flow. These pressure fluctuations from upstream to downstream cause the self-excited shock train oscillatory motion. Similarly, a separated shear layer flow with large-scale vortical structures propagates downstream from the leading edge of the cavity and collides with the trailing edge of the cavity. This results in an acoustic disturbance that travels in the upstream direction.
  
 \begin{figure}
\includegraphics[width=1\linewidth]{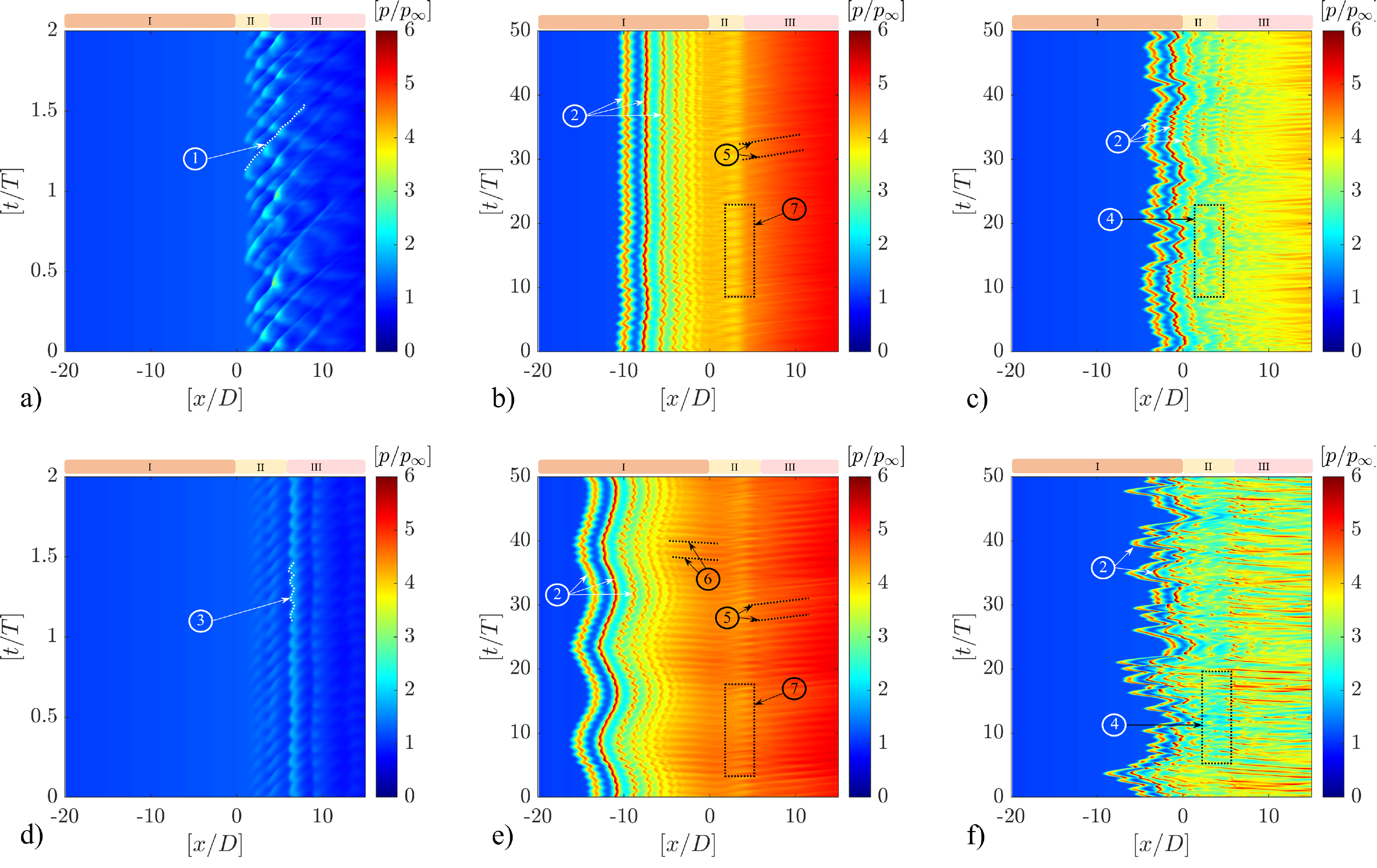}
  \caption{\label{fig:cl_press_xt} Non-dimensional static pressure [p/p$_{\infty}$] trace over time [t/T] along the duct centerline [i.e., y/D = 1.0].    [(a)-(c)] \& [(d)-(f)] shows cavity with front wall angle is 90$^o$ and 60$^o$. I) isolator section, II) cavity, III) diffuser section. Observed flow features are 1) traces of shear-induced weak shock wave, 2) successive shock from shock train, 3) traces of oblique shock wave from the front wall surface 4) shock train cavity shear layer interaction 5) disturbance propagating downstream in the diffuser section (right running wave) 6) disturbance propagating upstream in the isolator section (left running wave)  7) coalescence of right and left running wave.}
  \end{figure}
 Figure.\ref{fig:wall_press_xt} shows the static pressure distribution along the bottom wall traced over time [t/T], where T = 1ms. The x-axis is taken as the wall surface coordinate system (s/D) and the origin is set at the cavity leading edge point. The space-time contour shows the propagation of the pressure disturbance through the slope lines labeled (1,2,5,7) in Fig.\ref{fig:wall_press_xt}. The slope not only shows the disturbance propagation direction but also indicates the global propagation velocity. Here the global represents either for the total length of the duct or the entire flow field domain. Along the flow field, there may be a phase change in the disturbance propagation direction which can be seen in the subsequent section from the two-point cross-correlation method. The comparison between the different cases clearly shows the oscillatory motion of the cavity and the shock train. For the case without a shock train (see Fig.\ref{fig:wall_press_xt}(a) \& (d)), only the time period of [t/T] = 2 is shown to indicate various flow features in detail. This also conveys the frequency of self-sustained oscillation is larger in the cavity. Figure.\ref{fig:wall_press_xt}(b) and (e) show the sinusoidal motion of the shock train referred to as self-excited oscillation. In the case where the front wall of the cavity has an angle of 60$^o$ (case 2b), a large amplitude of the shock train oscillation is observed. It can also be seen that the effects of the shear flow on the shock train are smaller in cases 1a and 2a (i.e., p$_b$/p$_{\infty}$ = 6.0) than in cases 1b and 2b (i.e., p$_b$/p$_{\infty}$ = 5.0). At the same time, a strong influence of the diffuser section is observed in cases 1b and 2b due to the separated boundary layer flow. The static pressure along the centerline (y/D = 1.0 and -20 $\leq$ x/D $\leq$ 15) is probed and a space-time contour (x-t) similar to the wall static pressure data is constructed and shown in Fig.\ref{fig:cl_press_xt}. Figure.\ref{fig:cl_press_xt}(b,c,e,f) clearly shows successive shocks oscillating almost in phase with each other. Due to the interaction with a shear layer, the successive shocks tend to smear out and form a strong compression shock at the trailing edge of the cavity. This is also confirmed by Mach [M] contour and the vortex contour [$\omega$ D/u$_{\infty}$] (see Fig.\ref{fig:mach_instant} and \ref{fig:vort_instant}. Moreover, the shock train oscillatory fashion is erratic in the case of the back pressure ratio [p$_b$/p$_{\infty}$ = 5.0].

  \begin{figure}{}
\includegraphics[width=0.5\linewidth]{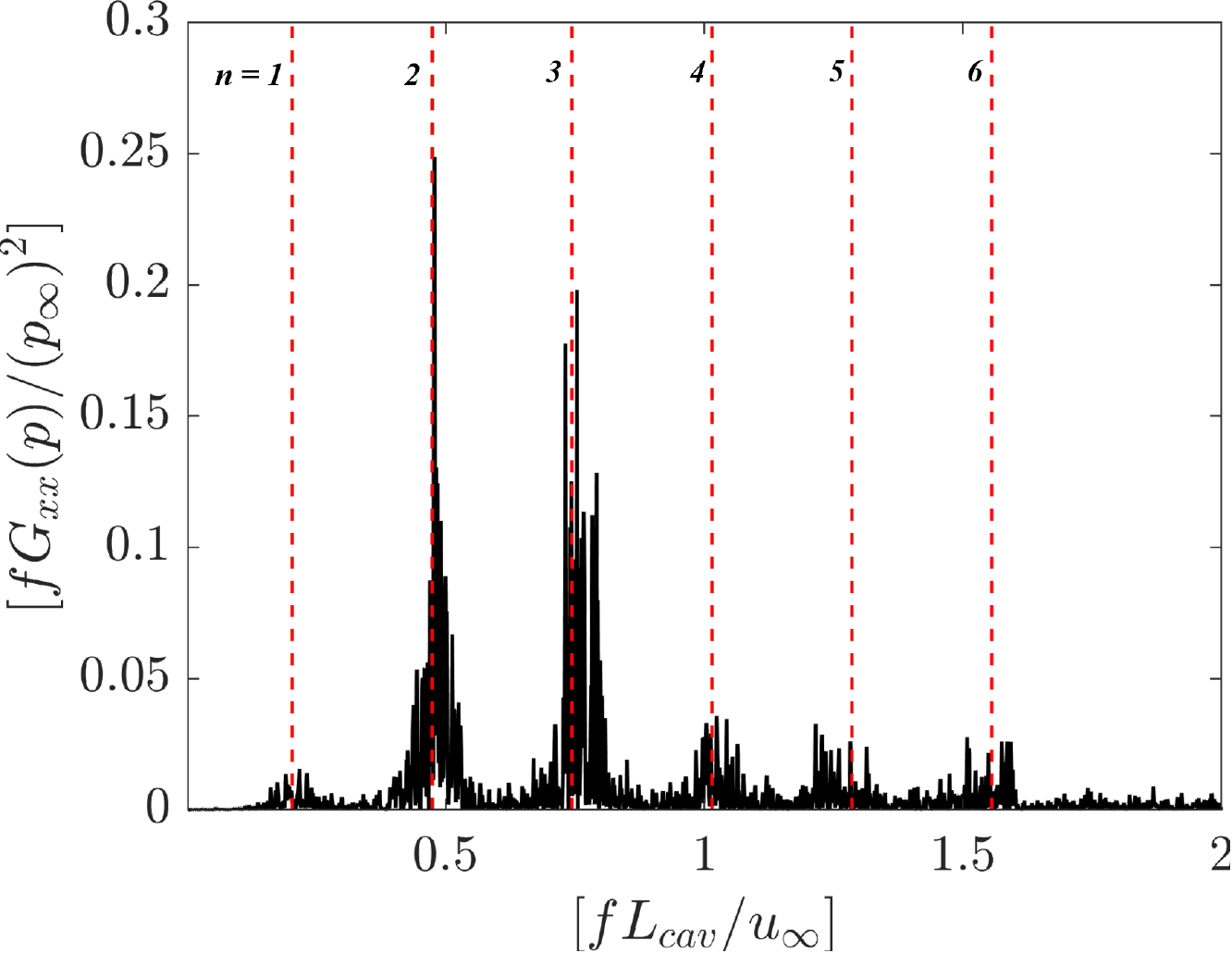}
  \caption{\label{fig:cav_0_psd} Non-dimensional power spectrum [fG$_{xx}$(p)/p$_{\infty}^2$] contour from centerline static pressure distribution (p/p$_{\infty}$). The dominant frequency from Rossiter's empirical relation is shown with the dashed red line.}
  \end{figure}

\begin{figure}
\includegraphics[width=1\linewidth]{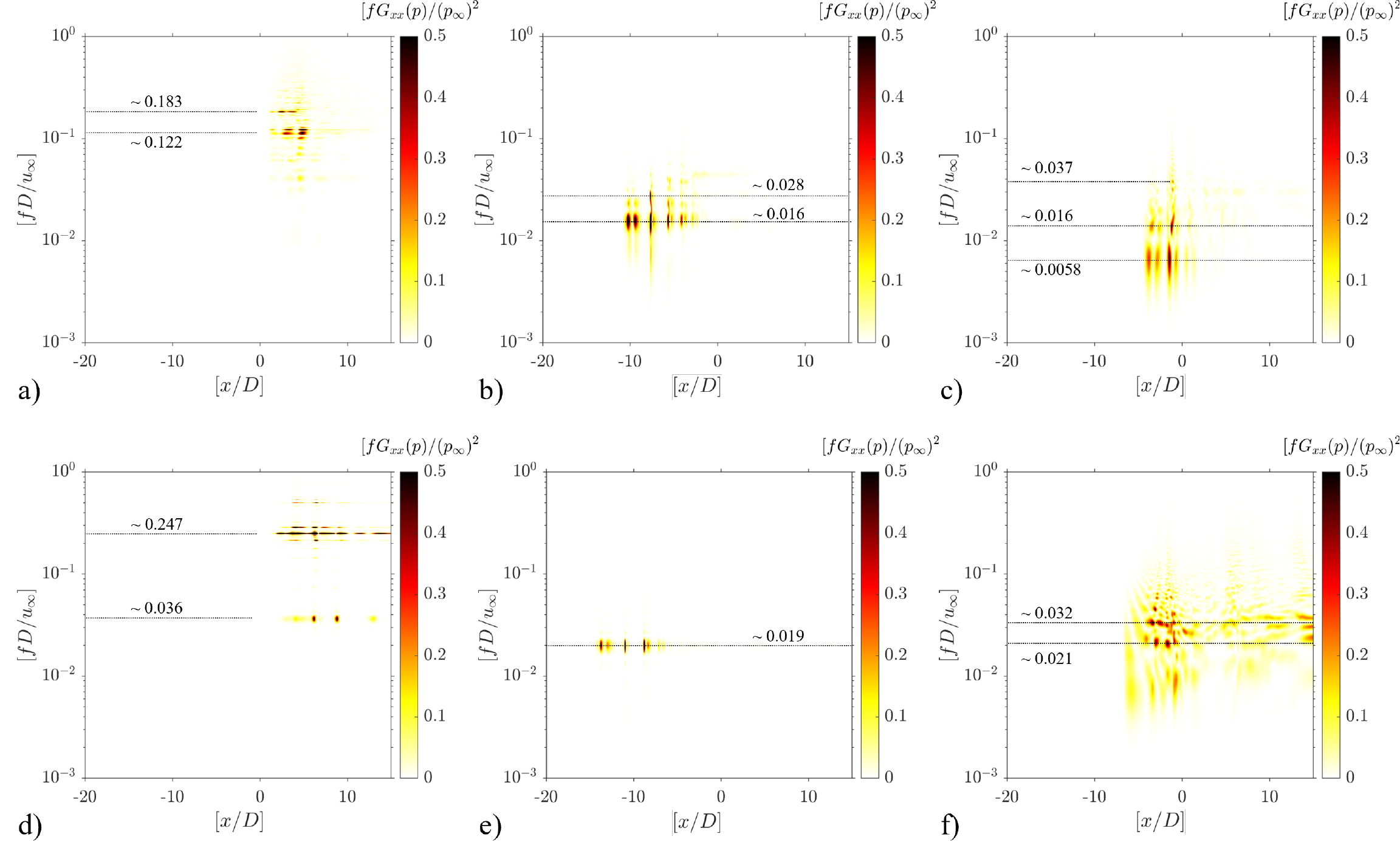}
  \caption{\label{fig:cl_psd} Non-dimensional power spectrum [fG$_{xx}$(p)/p$_{\infty}^2$] contour from centerline static pressure distribution (p/p$_{\infty}$). The dominant frequency is marked with the dashed line showing   }
  \end{figure}

   \begin{table}
  \caption{\label{tab:cav_0_psd} Comparison of cavity oscillation frequency (case 1) with Rossiter's empirical formula (Eq.\ref{eq:ross})}
  \begin{ruledtabular}
  \begin{tabular}{l c c }
  mode number 	& Present CFD mode        & Rossiter's mode          \\
  (n)             & \textit{fL$_{cav}$/u$_{\infty}$} & \textit{fL$_{cav}$/u$_{\infty}$}  \\
  \hline
  2  &  0.4780  &  0.4735       \\
  3  &  0.7536  &  0.7441  	  \\
  \end{tabular}
  \end{ruledtabular}
  \end{table}
  To analyze the oscillatory and dominant flow mechanism in detail, a spectral analysis using the extracted static pressure data was performed at different locations in the computational domain. Performing Fast Fourier Transformation (FFT) of the pressure signal using equation (Eq.\ref{eq:psd}) yields the power spectral density (PSD) \cite{ram_2022}.

 \begin{equation}
  \label{eq:psd}
  G_{xx}(p)=\lim\limits_{\Gamma\to\infty} \frac{1}{\Gamma} |F_\Gamma(p(t))|^2 
  \end{equation}
  where G$_{xx}$(p) is the power spectrum (unit: pa$^2$/Hz), the total duration of data acquisition is given by $\Gamma$, and \textit{|F$_\Gamma$(p(t))|$^2$} is the fast Fourier transform of the given pressure signal. The power spectra for case 1 from the static pressure data (p/p$_{\infty}$) data monitored at the center of the bottom wall of the cavity (S$_{17}$, at [x/D] = 2.0 and [y/D] = -1.0) are shown in Fig.\ref{fig:cav_0_psd} along with the Rossiter's mode (from Eq.\ref{eq:ross}). The second and third modes are dominant and are relatively close to the empirical Rossiter frequency shown in Table.\ref{tab:cav_0_psd} and other modes (1,4,5,6) appear to be split into several low-amplitude spectra. With the spatiotemporal power spectrum, it becomes easier to visualize the peak frequency range between the different cases. The two-dimensional power spectrum from the available static pressure distribution in the centerline is shown in Fig.\ref{fig:cl_psd}. The x-axis is the dimensionless surface length (s/D), the y-axis is the dimensionless frequency (\textit{fD/u$_{\infty}$}) (i.e., the Strouhal number), and the contour values correspond to the frequency multiplied power spectrum [G$_{xx}$(p)] non-dimensionalized by the square of freestream static pressure (p$_{\infty}$ = 28 kPa). The peak frequencies are marked and indicated with a dashed line. Fig.\ref{fig:cl_psd}(a and d) show the PSD without shock train for the 90$^o$ and 60$^o$ cavity front wall angle cases respectively. The peak frequency (fD/u$_{\infty}$) without shock train falls in the range of 0.122 $\sim$ 0.183 for case 1 and 0.036 $\sim$ 0.247 for case 2. The corresponding dimensional frequency [f] is $>$ 5kHz, which falls under the category of high-frequency oscillations. These high-frequency oscillations are prone to combustion instabilities and buzz flow in the scramjet engine. It appeared that the peak frequency was oriented towards the trailing edge of the cavity, and a reduction in peak frequency [fD/u$_{\infty}$ $\sim$ 0.036] was observed for the cavity front wall angle of 60$^o$ (as marked in Fig.\ref{fig:cl_psd}(d)). For the cases with a shock train, the peak frequencies are shown near the shock train locations. The high oscillatory frequency of the cavity flow seems to be naturally controlled by the low-frequency oscillation of the shock train. A reduction of more than 50\% in cavity oscillation is observed. This in turn provides a longer time for the fuel-oxidizer mixing and flame stabilization process. When the shock train is present near the cavity, a strong interaction takes place between the cavity shear flow and shock train, which is shown earlier in the x-t contour. The power spectrum for cases 1b and 2b shows several peak frequency components due to the shock train and shear layer interaction. Several patches of frequency components in the diffuser region (see Fig.\ref{fig:cl_psd}(f)) clearly indicate that the downstream separated flow oscillates in phase with the shock train or vice-versa.           

\subsection{Disturbance propagation mechanism}
\label{sec:4c}  
\begin{figure}
    \subfigure[]{\label{fig:cav_0_ccr_plot}\includegraphics[width=0.5\linewidth]{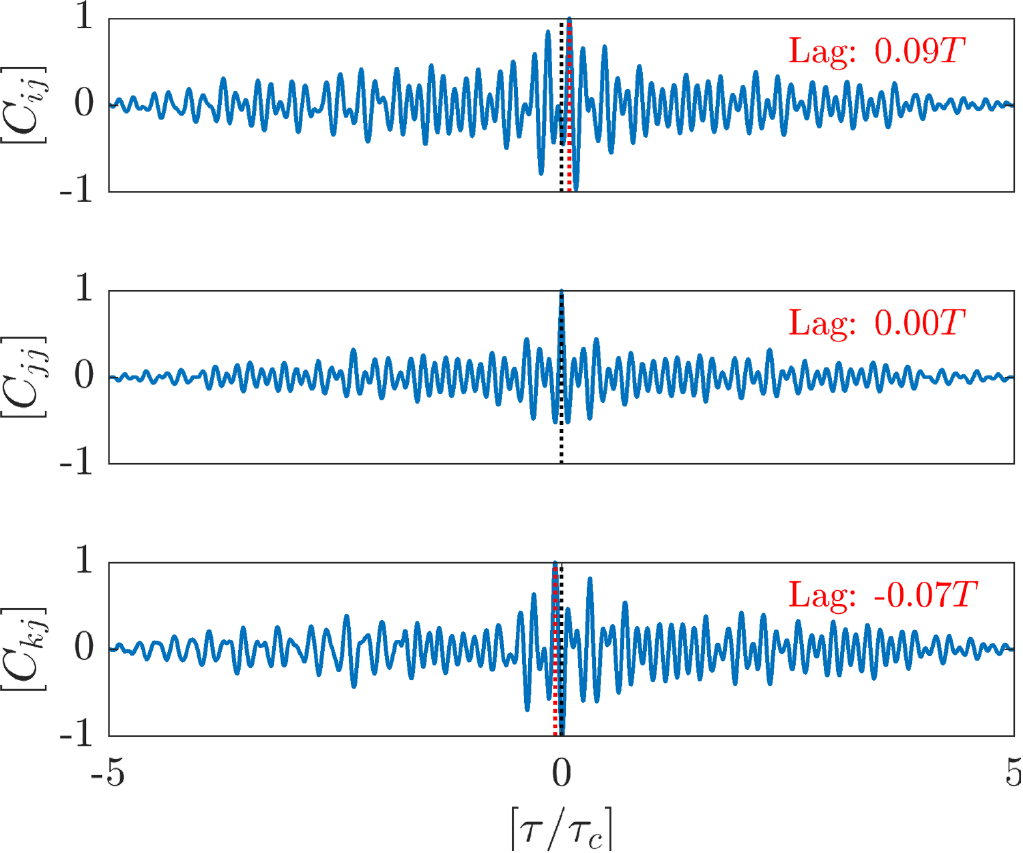}}
    \hspace{0.5cm}
    \subfigure[]{\label{fig:prop_mech}\includegraphics[width=0.3\linewidth]{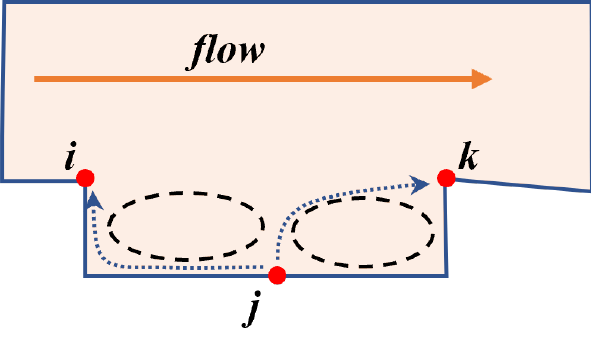}}
    \caption{\label{fig:cross_cor} a) Two-point cross-correlation coefficient between ij, jj and jk points b) schematic of disturbance propagation direction}
  \end{figure}  
  The acoustic disturbance in the scramjet isolator can propagate either upstream or downstream at a finite velocity, which can initiate a local oscillatory flow. Such disturbances can only propagate through the subsonic boundary layer flows\cite{ram_2022}. Although the x-t contour shows the convection direction of the disturbance (1,2,5,7), as marked in Fig.\ref{fig:wall_press_xt}, it is possible to estimate the propagation velocity of the pressure disturbance from the slope. But a local disturbance that also causes the self-excited oscillatory motion cannot be predicted. To understand the propagation mechanism of such a disturbance, the extracted static wall pressure data can be processed for a two-point cross-correlation (C$_{ij}$) given by equation (Eq.\ref{eq:ccr}),
  
  \begin{figure*}
\includegraphics[width=1\linewidth]{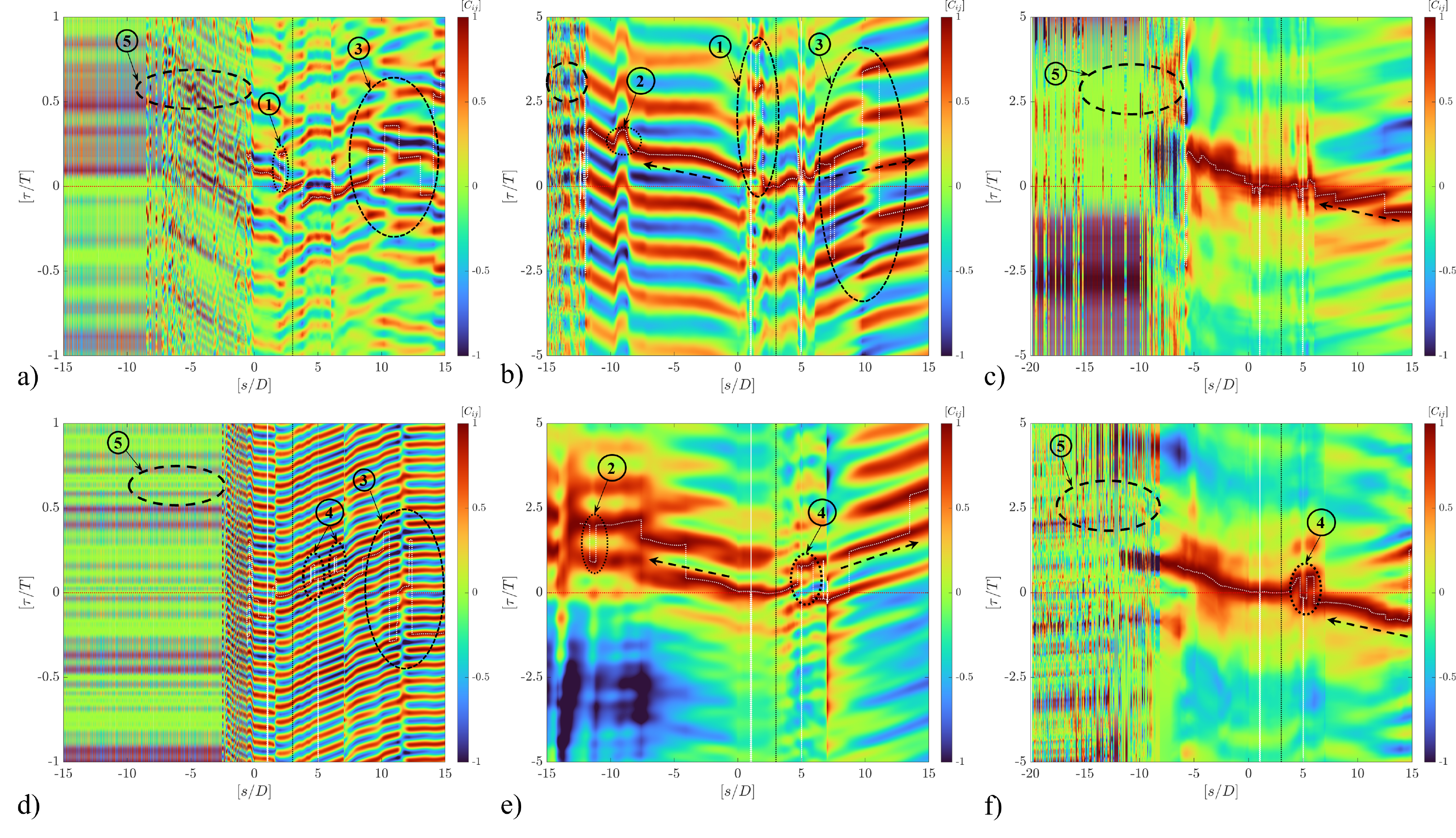}
  \caption{\label{fig:dist_prop} Spatiotemporal cross-correlation coefficient contour analyzed from bottom wall static pressure distribution (p/p$_{\infty}$). The maximum cross-correlation coefficient at each spatial node is shown as a white dotted line.}
  \end{figure*}
\begin{equation}\label{eq:ccr}
  C_{ij}(\tau) = \frac{\lim\limits_{\zeta\to\infty} \frac{1}{\zeta} \int_{0}^{\zeta} p_{i}^{'}(t) p_{j}^{'}(t+\tau)dt} {\left(\lim\limits_{\zeta\to\infty} \frac{1}{\zeta}\int_{0}^{\zeta} p_{i}^{'2}(t) dt \right)^{1/2} \left(\lim\limits_{\zeta\to\infty} \frac{1}{\zeta}\int_{0}^{\zeta} p_{j}^{'2}(t) dt \right)^{1/2}}
\end{equation}

  where '\textit{i}' is the location cross-correlated with the reference location ('\textit{j}'), $\tau$ is the time delay between (\textit{i-j}), $\zeta$ is the total duration of data sampling time, p$_i$ and p$_j$ are the corresponding pressure-time histories at i and j, respectively. In the present computational analysis, the cross-correlation coefficients (C$_{ij}$) are calculated with respect to the reference point (\textit{j}) and considered at the mid-location of the cavity bottom wall (i.e., [s/D] = 3.0). Figure.\ref{fig:cav_0_ccr_plot} shows an example of the cross-correlation between the leading-edge point and the trailing edge point with respect to the mid-point of the cavity bottom wall. The delay time between two sampling points directly indicates the propagation direction. In the present analysis, a positive time delay means that the disturbance propagates upstream, and a negative time delay means that the disturbance propagates downstream. For the same point, the correlation of a signal is an autocorrelation and shows a time delay of zero. Plotting two-dimensional cross-correlation coefficients at each point along the bottom wall can clearly see the phase shift of the local disturbance convection direction. Fig.\ref{fig:dist_prop} shows the spatiotemporal distribution of the cross-correlation coefficient with respect to the cavity bottom wall (at [s/D] = 3.0). The y-axis shows the time lag of each spatial coordinate along the bottom wall with the reference point. The white dotted line indicates the time lag [$\tau$/T] of the peak cross-correlation coefficient [C$_{ij}$]. As mentioned earlier, the positive value of the time lag [$\tau$/T] indicates that the disturbance propagates in the upstream direction and a negative time lag [$\tau$/T] represents the disturbance propagates downstream. The black dashed vertical line (at [s/D] =3.0) indicates the reference location from which the cross-correlation coefficient is calculated. It can be seen that the propagation of the disturbance is not a linear smooth change either upstream or downstream. Fig.\ref{fig:dist_prop} clearly shows a local phase shift in the propagation direction of the disturbance. This is due to the interaction of counter-rotating vortices in the separated boundary layer, recirculating vortices in the cavity approaching the corner point, and the impingement of the shear layer on the front wall or trailing edge point. At these points, a small-scale vortex formed, triggering a local self-sustaining oscillation. The present cases also show a small-scale disturbance marked as 5 in Fig.\ref{fig:dist_prop}. This can be solved using higher-order numerical techniques such as direct numerical simulation (DNS). The averaged propagation of the disturbance is indicated by a black dashed arrow line, which may be the slope line of the x-t contour. A sudden rise and dip in the time delay are due to the separation and reattachment of the boundary layer. Fig.\ref{fig:dist_prop}(a,b,d,e) shows the propagation of the disturbance in both directions with respect to the reference point. In the presence of a shock near the cavity (cases 1b and 2b), the disturbance propagates only in the upstream direction. This is due to the strong influence of the separated flow in the diffuser region.
  
\subsection{Modal analysis}
\label{sec:4d}
 \begin{figure}
  \includegraphics[width=0.6\linewidth]{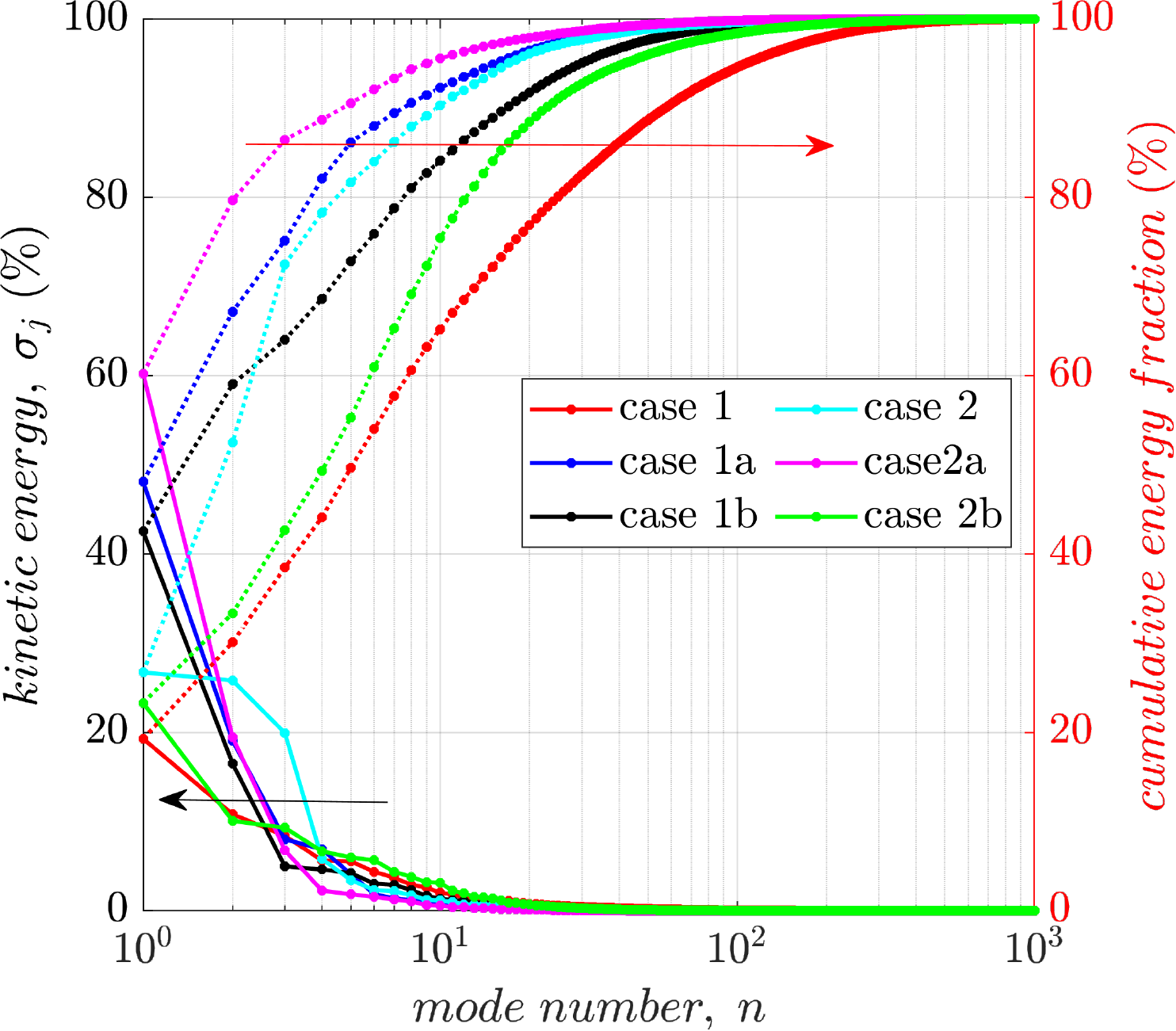}
  \caption{\label{fig:pod_tke} Comparison of kinetic energy spectra [$\sigma_j$] and cumulative energy fraction for different cases with respect to POD modes. ( for cases details refer Table.\ref{tab:cases} }
  \end{figure}

  Proper orthogonal decomposition (POD) is a technique for determining the orthonormal basis from a set of input data. Further, DMD analysis can provide sufficient information about dominant spatiotemporal modes between the shock train and shear layer interaction based on their frequency spectra \cite{schmid_2010}. In the present analysis, the spatial POD and DMD algorithms of Tu .et.al \cite{tu_2013} are used to identify the coherent structures based on their energy and frequency content. For the analysis, a set of velocity fields extracted from the fluid domain at each time step is reconstructed with an equidistance matrix of 1001 $\times$ 101 for a physical time of 50 ms. Later it is stacked in as a data matrix of 1000 snapshots with $\Delta$t = 5e$^{-5}$ s. The eigenvalues obtained from the POD analysis are directly proportional to the total kinetic energy (TKE) of each POD mode. Then the eigenvalues are sorted in the order of highest to lowest to ensure that the most important modes consist of large-scale flow structures. The calculated kinetic energy associated with each mode number (n) is plotted in Fig.\ref{fig:pod_tke} for 1000 modes. It can be seen that the total kinetic energy is retained within the first 10 modes itself. For cases (1, 2, and 2b) at least five main modes are required to associate more than 50\% of the total kinetic energy (Fig.\ref{fig:pod_tke}). For case 2a, the flow field within the first five modes can be convective because it contains more than 90\% of the TKE. At the same time, more than 100 modes are required to capture 90\% of the TKE for case 1.
  
  \begin{figure*}
\includegraphics[width=1\linewidth]{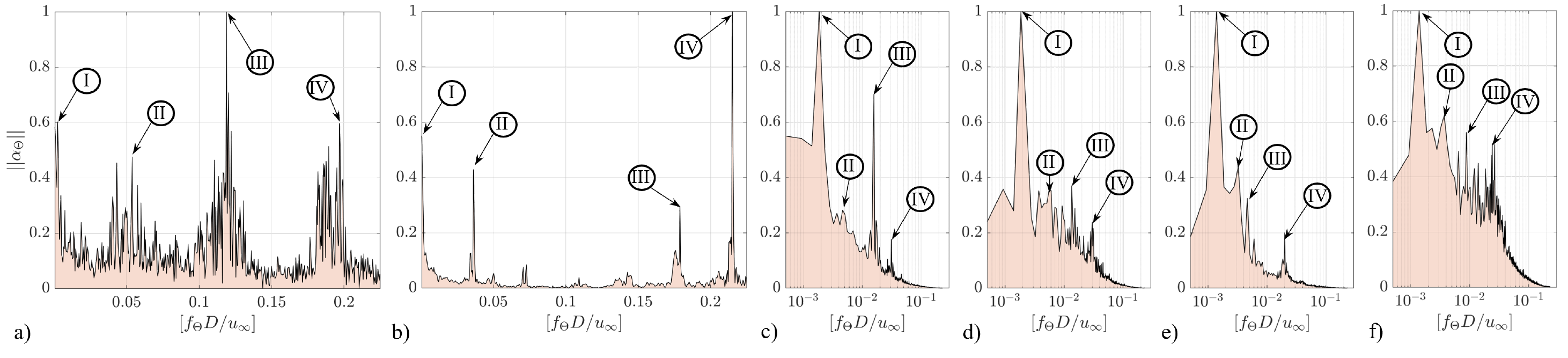}
  \caption{\label{fig:mode_freq} Normalized temporal spectra from DMD analysis (f$_\Theta$D/u$_{\infty}$ vs $||\alpha_{\Theta}||$) for different cases (in the order of 12,1a,2a,1b,2b refer Table.\ref{tab:cases}). Discrete peak frequencies marked as I, II, III, IV are taken for DMD modes visualization}
  \end{figure*}

  Fig.\ref{fig:mode_freq} shows normalized temporal spectra from DMD analysis (f$_\Theta$D/u$_{\infty}$ vs $||\alpha_{\Theta}||$) for different cases (in the order 1,2,1a,2a,1b,2b). In the cases without the shock train, the cavity flow experiences high-frequency oscillation, and in the other cases with the shock train shows the DMD spectrum falls in the frequency range [f$_\Theta$D/u$_{\infty}$] of 0.05. In the 90$^o$ cavity case, the dominant frequency corresponding to n = 1,2,3,4 is observed and is close to the Rossiter mode, as shown in the power spectral density plot (Fig.\ref{fig:cav_0_psd}). In the case of the 60$^o$ cavity, the peak frequencies (II and IV) marked in Fig.\ref{fig:mode_freq}(b) is clearly visible in the spectral contour (fD/u$_{\infty}$ $\sim$ 0.036 and 0.247). The mode number (n) corresponding to 1, 5, and 6 is dominant. The discrete peak for frequency [f$_\Theta$D/u$_{\infty}$] $<$ 0.05 is due to the oblique shock wave formed at the front wall, and higher frequencies are due to the shear layer interaction. For cases (1a, 2a, 1b, and 2b), the most dominant frequency was observed near $\sim$ 0.0018. The corresponding frequency is in the limit of low frequencies caused by shock train oscillation (f $\sim$ 200 Hz) \cite{geerts_2016,gnani_2016,ram_2021b}. In the case of the 90$^o$ cavity (case 1a), a peak frequency was observed near $\sim$ 0.028 (marked as III) in Fig.\ref{fig:mode_freq}(c), which also occurs in all other cases (marked as IV) with shock train flows.
  
  \begin{figure*}
    \subfigure[]{\label{fig:cavity_0}\includegraphics[width=0.45\linewidth]{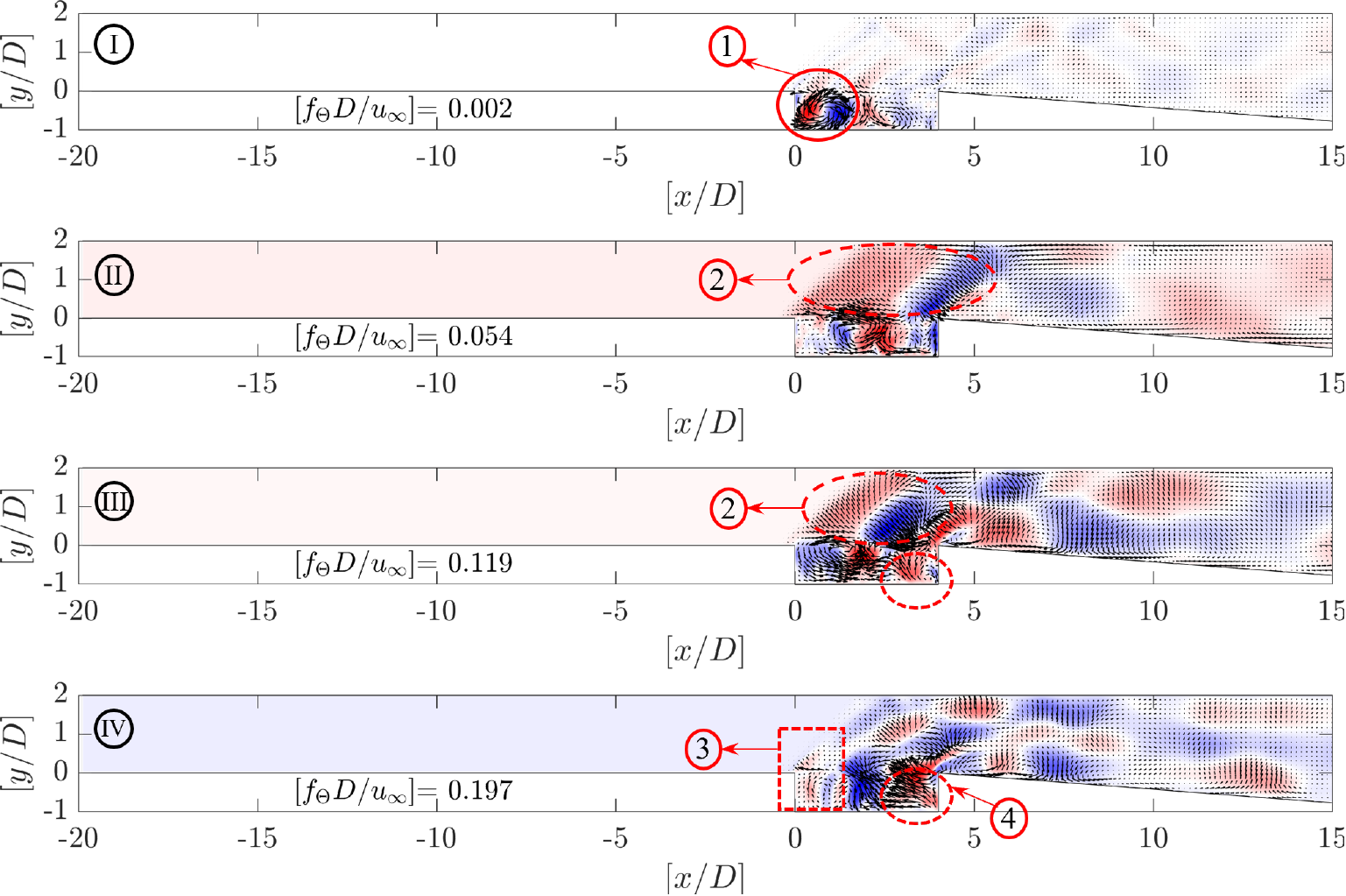}}
    \hspace{0.5cm}
    \subfigure[]{\label{fig:cavity_60}\includegraphics[width=0.45\linewidth]{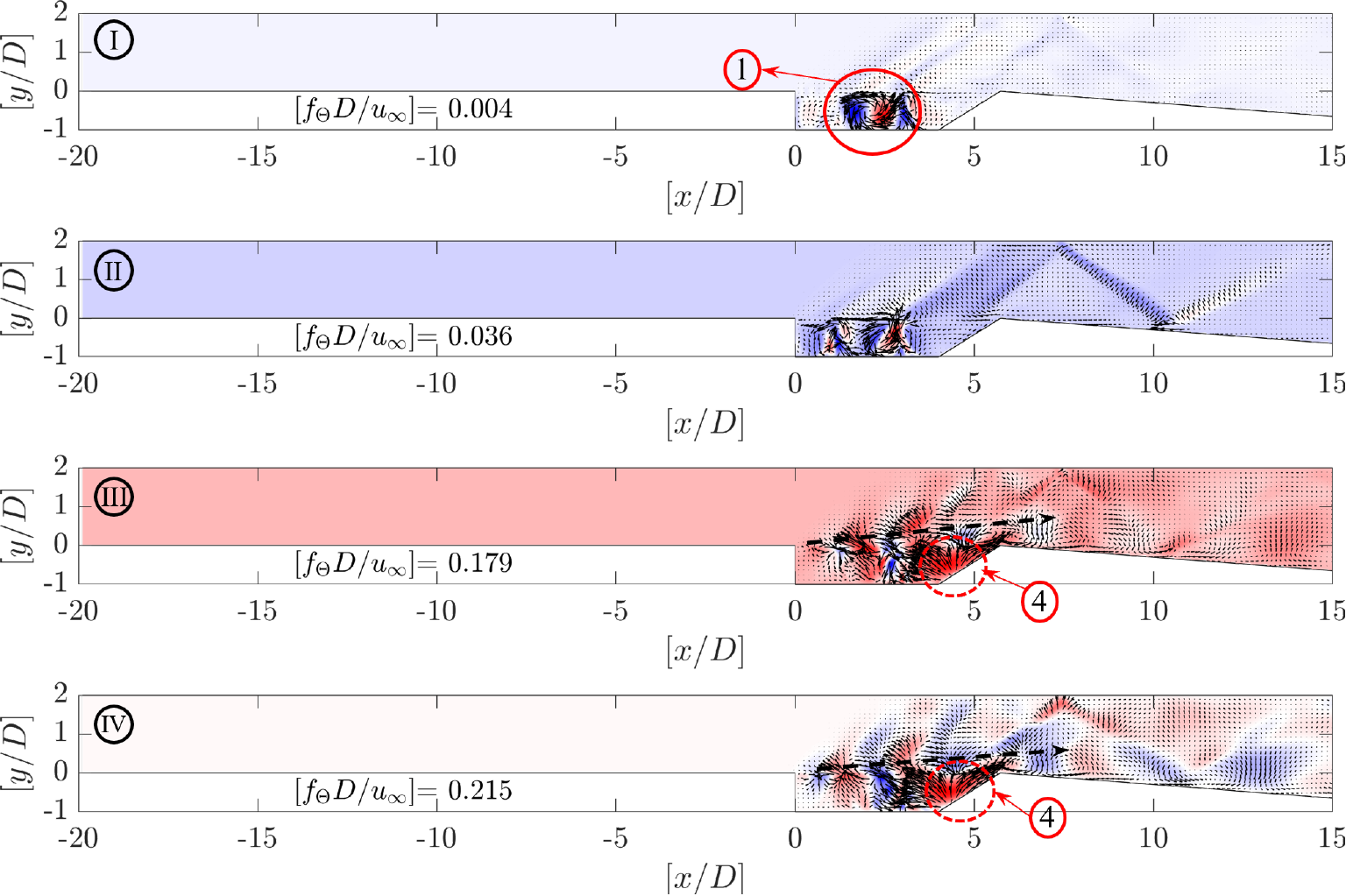}}
    \subfigure[]{\label{fig:ram_0}\includegraphics[width=0.45\linewidth]{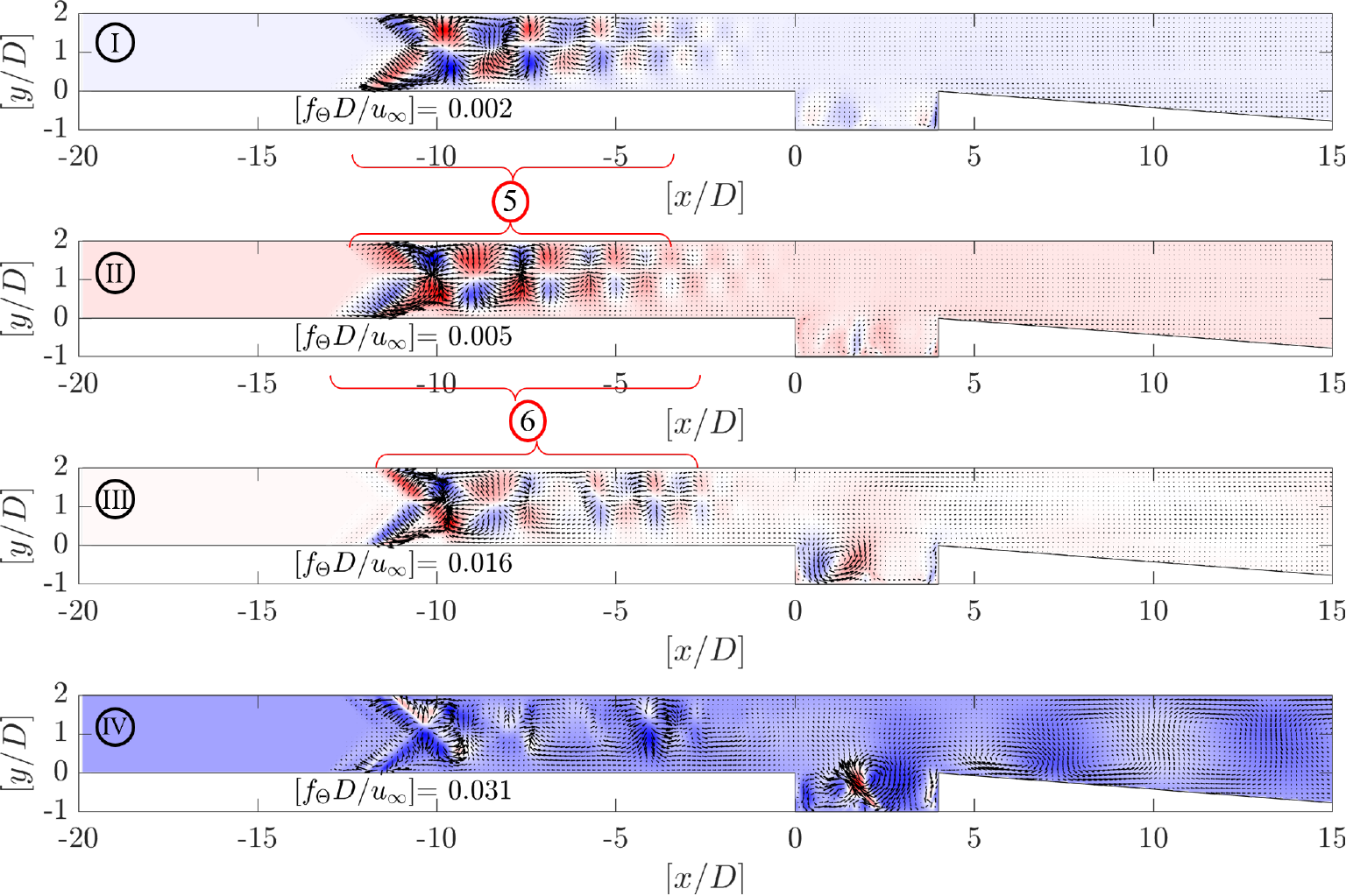}}
    \hspace{0.5cm}
    \subfigure[]{\label{fig:ram_60}\includegraphics[width=0.45\linewidth]{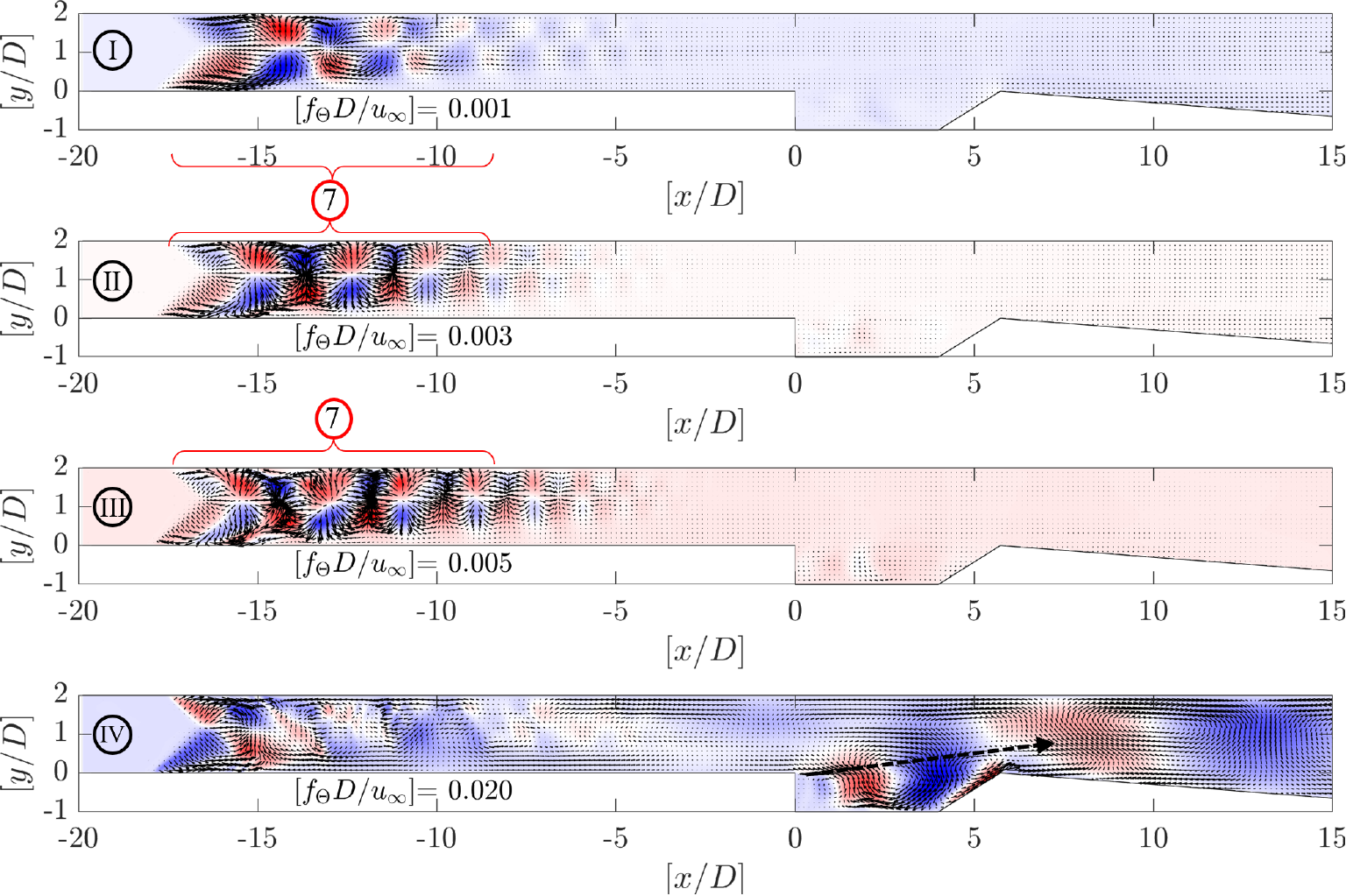}}
    \subfigure[]{\label{fig:scram_0}\includegraphics[width=0.45\linewidth]{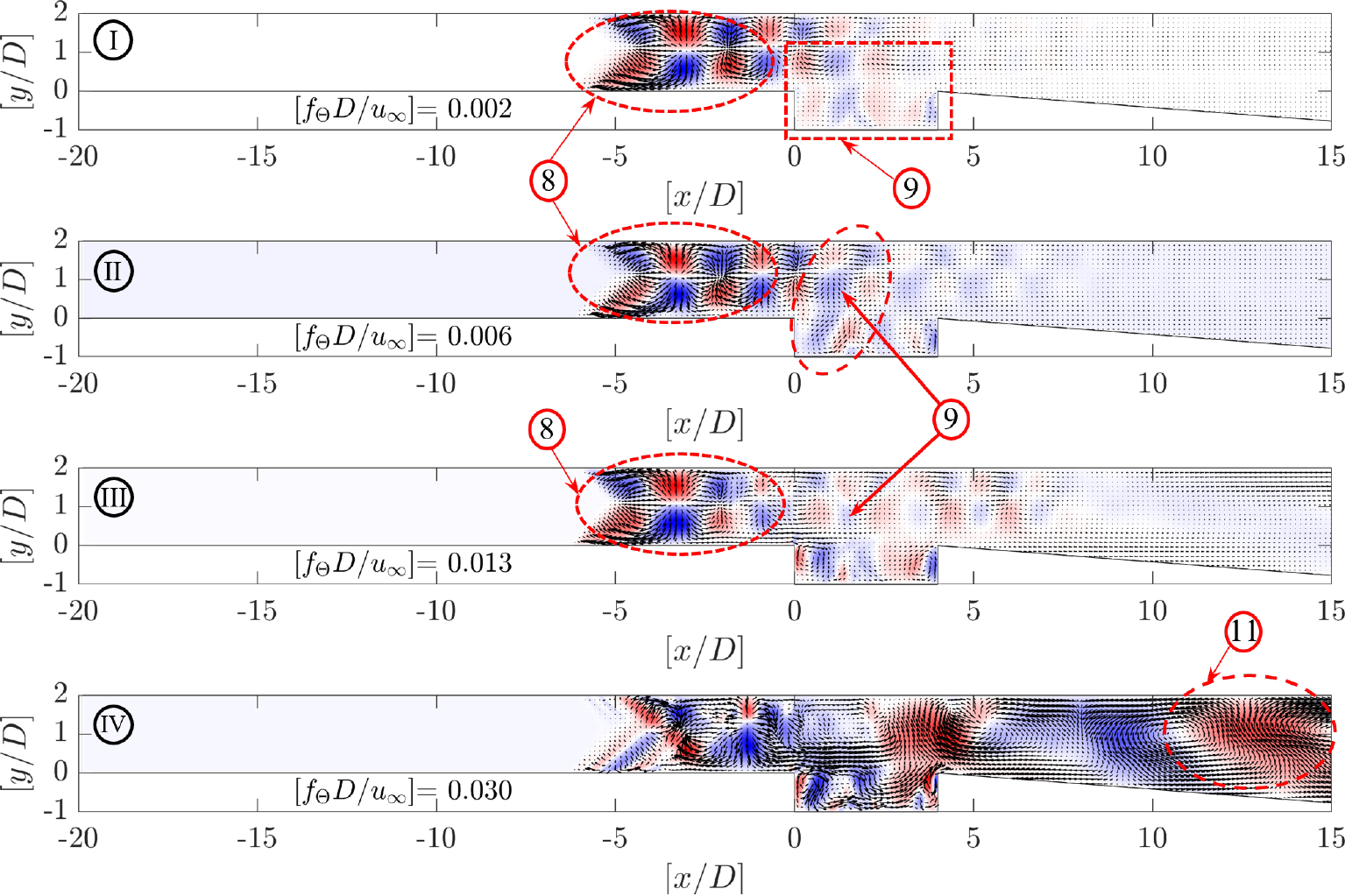}}
    \hspace{0.5cm}
    \subfigure[]{\label{fig:scram_60}\includegraphics[width=0.45\linewidth]{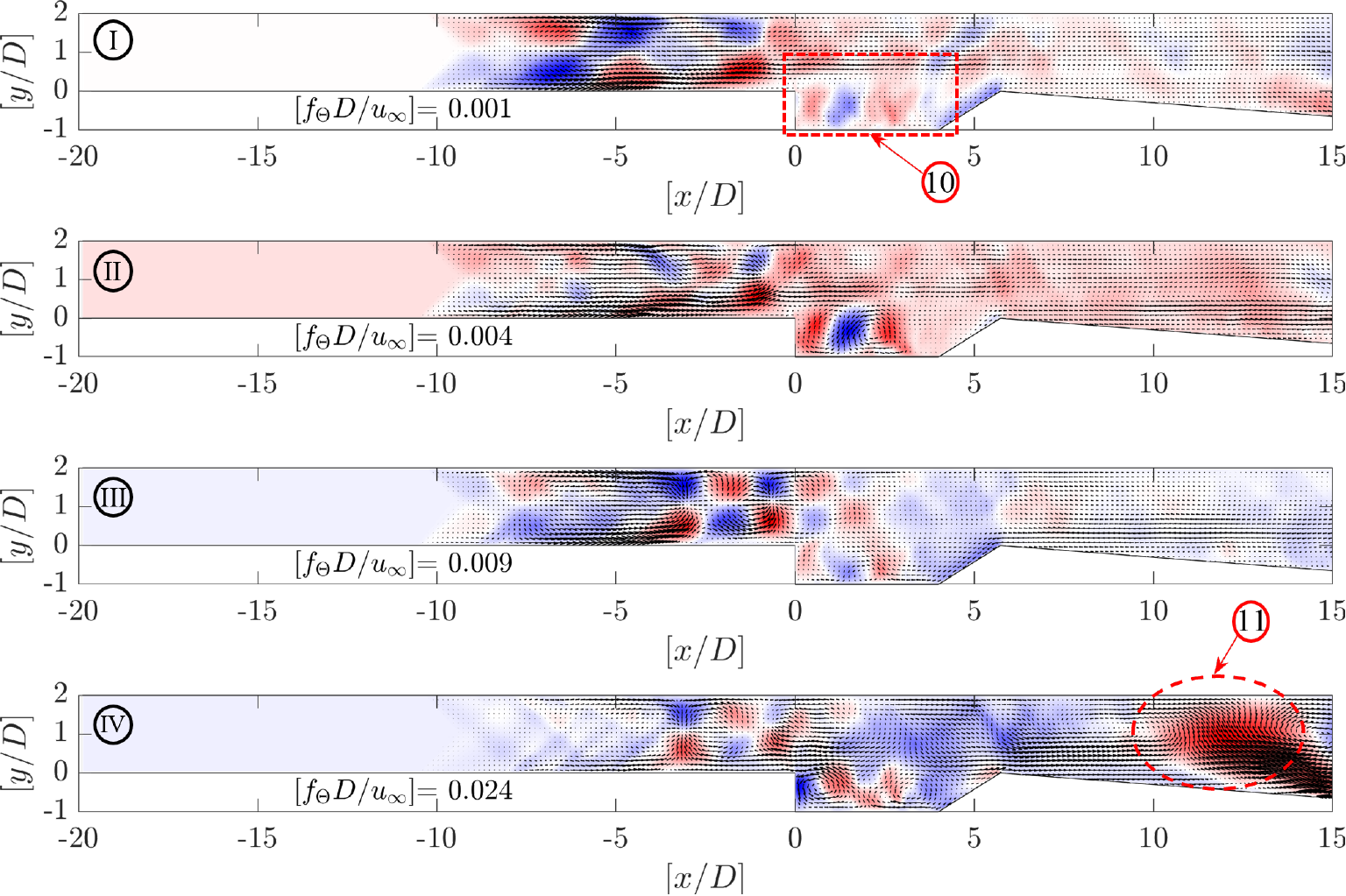}}
    \caption{\label{fig:DMD_modes} Normalized spatial DMD modes [$\Theta$] corresponding to the velocity field. Marked labels are given in the description.}
  \end{figure*}  

  Figure.\ref{fig:DMD_modes} shows the DMD modes in the flow field from discrete DMD peaks (labeled as I, II, III, IV in Fig.\ref{fig:mode_freq}). The corresponding frequency value is noted, and the contour values are scaled between -1.0 and 1.0, along with the quiver plot of the velocity field. For cases 1 and 2, mode (I) shows a strong vortical structure (marked as 1) inside the cavity. These vortical structures are very similar to the secondary vortices inside the cavity (see Fig.\ref{fig:mach_vorticity}). The most dominant mode seems to emanate from the secondary vortices. In case 1, modes II and III show the coherent structures (marked as 2) originating from the shear-induced weak shocks. In the fourth mode (IV), the coherent structures (marked as 2) convect downstream to the front wall (marked as 4) of the cavity and form a new structure (marked as 3) near the rear wall of the cavity. Similarly, in case 2, the dominant structure appears to convect downstream (Fig.\ref{fig:DMD_modes}(b)). An oblique and reflected shock structure is also clearly visible in all four modes presented here (I $\sim$ IV), and in mode (IV) the shear layer interacts with the weak compression waves downstream of the cavity. In cases 1a and 2a, the discrete dominant modes are due to the shock train flow field. The fourth mode (IV) shows a weak coherent structure in the diffuser section. In the case of a 90$^o$ cavity, the modes (I and II) appear to be opposite in nature, while the coherent structure of the shock train in the modes (II and III) is exactly the same. The structure of the shock train (marked as 7) appears to be the same except for its length in case 2a (Fig.\ref{fig:DMD_modes}(d)). When the shock train is present close to the cavity, the interaction with the shear layer is predominant. Case 1b shows that the coherent structure from the shock train strongly correlates with the cavity flow field (marked as 9) and the pairing of coherent structures between the shock train and cavity recirculation flow is observed. These coherent structures moved downstream with subsequent dominant modes. In the case of the 60$^o$ cavity, the coherent structures are smeared out due to the large amplitude of shock train oscillation. In both case 1b and case 2b, large vortical structures can be seen and convecting towards the upstream direction. These structures strongly induce the oscillatory motion of the shock train.

\subsection{Limitations and Future Scope}
\label{sec:4e}
  In the present study, a two-dimensional flow field in a scramjet isolator with a cavity was investigated. The findings are related to the interaction between the shock train- cavity shear layer, where the flow is mainly dominated by longitudinal oscillatory motions. In an actual flow field, the effects of sidewalls and flow separation from corners can cause lateral oscillations. The present results may change relatively in the dominant spectral values due to the inclusion of sidewalls, fuel injection, combustion, and flame propagation. Therefore, it is necessary to extend such studies with high-fidelity calculations along with parametric studies that include different cavity sizes, free stream Mach number, and fuel injection. Such studies will be conducted to characterize the flow field and left as a future research scope.     
\section{CONCLUSIONS}
\label{sec:5}
  The interaction of the shock train with the cavity shear layer in a scramjet isolator is analyzed numerically using a commercial flow solver. To resolve the unsteady dynamics of the oscillating flow field, the hybrid RANS/LES-based methodology is applied. The effect of the position of the shock train and the angle of the cavity front wall is investigated. The position of the shock train in the isolator is achieved by changing the back pressure ratio (p$_b$/p$_{\infty}$ = 5.0 and 6.0). The summary of the main results is presented below.
  
\begin{enumerate}
   
    \item The time-averaged static pressure distribution along the centerline (y/D = 1.0) shows that the maximum pressure increase due to the first shock is less than that of the second shock. The structure of the first shock appears to be oblique because the freestream Mach number range falls in the category of transition between normal and oblique shock train. The transition from normal to oblique shock train may occur due to the duct blockage ratio.
    \item In the space-time contour, the propagation of the pressure perturbations can be clearly seen. The pressure fluctuations caused by the cavity propagate in both upstream and downstream directions. The successive shock structure and its oscillatory motion are distinctly visible from the spatiotemporal pressure distribution along the centerline. The strong interaction between the shock train and the cavity shear layer leads to an irregular oscillation at a backpressure ratio (p$_b$/p$_{\infty}$) of 5.0. At a backpressure ratio of 6.0, however, the oscillation of the shock train is sinusoidal.
    \item The power spectrum of the cavity flow without shock train case (p$_b$/p$_{\infty}$ = 0.7) shows a high-frequency oscillation. The peak frequencies are compared with Rossiter's mode for the angle ($\theta$) of the cavity front wall of 90$^o$. It is possible to identify the first five modes, with modes (n = 2 and 3) being dominant and correlating well with Rossiter's mode. The spatiotemporal power spectrum clearly shows that the low-frequency high-amplitude shock train flow is dominant. The frequency of cavity flow oscillation is reduced by more than 50\%. The interaction between the shock train and cavity flow tend to reduces the cavity oscillation frequency which can help in fuel-oxidizer mixing and flame stabilization. 
    \item Through the spatiotemporal cross-correlation coefficient, the time delay between the pressure disturbance shows a local phase shift in the disturbance propagation direction. These local disturbances initiate fluid flow instabilities.
    \item Modal analysis is carried out by using proper orthogonal decomposition and dynamic mode decomposition (DMD) methods. The cumulative energy fraction shows that more than 100 modes are required to capture 90\% of the dominant flow structures for case 1. In cases (1a and 2a), where the shock train occurs well upstream of the cavity, only 10 modes are required. In cases (1b and 2b), 25 modes are required to capture 90\% of the dominant flow field structures. The cases with large modes requirement are due to the flow consisting of various ranges of turbulent length scales.
    \item Using dynamic mode decomposition (DMD), the spectra of several discrete dominant modes are analyzed. DMD analysis of the case with the back pressure ratio of 0.7, shows the secondary vortices inside the cavity are dominant. While the back pressure ratio of 5.0 and 6.0, the peak dominant mode is due to the shock train flow field. A strong similarity and pairing of coherent structures exist between the shock train and the cavity flow for the back pressure ratio of 5.0.

\end{enumerate}

\section*{ACKNOWLEDGMENTS}
  This work was supported by the National Research Foundation of Korea (NRF) grant funded by the Korean government (MSIT) (No. NRF-2020R1F1A1075716).
  The first author would like to thank Dr.S.K.Karthick and Jintu K James for providing insight on computations and post-processing techniques.    

\section*{DATA AVAILABILITY STATEMENT}
The data that supports the findings of this study are available from the corresponding author upon reasonable request.

\nocite{*}
\bibliography{references2}

\end{document}